\documentclass[final,numbers,3p,times]{elsarticle}
\usepackage{amssymb}
\usepackage{amsthm}
\usepackage{amsmath}
\usepackage{bm}
\usepackage{graphicx}
\graphicspath{{Figures/}}		
\usepackage{tabularx}			
\newcolumntype{L}[1]{>{\hsize=#1\hsize\raggedright\arraybackslash}X}
\newcolumntype{R}[1]{>{\hsize=#1\hsize\raggedleft\arraybackslash}X}
\newcolumntype{C}[1]{>{\hsize=#1\hsize\centering\arraybackslash}X}

\usepackage{geometry}	
\usepackage{psfrag}

\usepackage[colorlinks=true]{hyperref}

\usepackage[resetlabels]{multibib}
\usepackage[utf8]{inputenc}
\usepackage{natbib}
\setcitestyle{square, comma, numbers,sort&compress, super}
\usepackage{soul}
\usepackage{xcolor}
\usepackage{color}

\usepackage{caption}
\usepackage{subcaption}

\journal{Journal}
\definecolor{lightgreen}{rgb}{.90,1,0.95}
\definecolor{lightblue}{rgb}{.90,0.95,1}

\begin{document}

\begin{frontmatter}

\title{Framework for a simulation-based aerodynamic shape optimization of bridge decks for different limit state phenomena}

\author[]{Tajammal Abbas\corref{cor1}${^{a}}$}
\ead{tajammalabbas@gmail.com}
\cortext[cor1]{Corresponding author,} 
\author[]{Igor Kavrakov${^a}$}
\author[]{Guido Morgenthal${^a}$}
\author[]{Tom Lahmer${^b}$}
\address[1]{Bauhaus University Weimar, Institute of Structural Engineering, Chair of Modelling and Simulation of Structures, \\Marienstra\ss e 13A, 99423 Weimar, Germany}
\address[2]{Bauhaus University Weimar, Institute of Structural Mechanics, Chair of Stochastics and Optimization, \\Marienstra\ss e 15, 99423 Weimar, Germany}

\begin{abstract}
Wind-induced response governs the design of the long-span bridges. 
The shape of the deck is one of the most important factors that not only affects the mechanical properties but greatly influences the aerodynamic performance of the bridge. 
An efficient framework is proposed to perform aerodynamic shape optimization of a bridge deck using Computational Fluid Dynamics (CFD) simulations and response surface strategies considering aeroelastic phenomena such as flutter, buffeting and Vortex-induced Vibrations (VIV). 
A parameterized cross section of the deck has been utilized ensuring required carriageway width and structural demand. 
The Particle swarm optimization algorithm has been employed which leads to an optimized shape that performs better as compared to the reference section. 
The presented aerodynamic shape optimization framework has a great potential to be used in the design of long-span bridges. 
\end{abstract}

\begin{keyword}
Aerodynamic optimization; long-span bridges; CFD; response surface; aeroelastic analysis.
\end{keyword}

\end{frontmatter}


\section{Introduction}
\label{Section:Introduction}
The number of long-span bridges built in the last few decades has increased considerably with a trend to have a slender and lighter deck. 
The flexibility and slenderness of these structures make them prone to wind actions which demand superior performance for serviceability and ultimate limit states. 
The evaluation of the aeroelastic behaviour of long-span bridges is vital for the design as they can develop significant vibrations and may lead to a catastrophic situation. 
The aerodynamic shape of the deck mainly influences the wind loads. 
This highlights the importance of choosing an appropriate deck shape for the design which is a challenging task. 
Conventional design approaches mainly depend on the wind tunnel tests where only a limited number of candidate geometries can be investigated to find the one with best performance. 
With the growth in the computational resources and development in the field of Computational Fluid Dynamics (CFD), numerical simulations offer a versatile and reliable method to assess wind effects on structures. 

The unconventional design by using optimization strategies has the capability to provide more efficient structures.  The optimization is commonly performed by minimizing the objective function that is encapsulated by minimizing the weight of the deck and satisfying the limit states for different phenomena. The geometry of the decks is generally parameterized so that a shape optimization, which falls into the class of non-linear and finite-dimensional optimization problems, needs to be solved. The optimization therefore offers a powerful technique to reduce wind effects on bridges. 
Adjusting the shape of the deck to satisfy the limit states would be therefore a viable solution and highlights its importance. 

The improvement in the aerodynamic behaviour due to shape change have been extensively reported for building structures. 
However, the utilization of optimization process in bridges has not attracted much attention. 

\citet{SimoesNegrao1994J} utilized entropy-based optimization algorithm for steel cable-stayed bridge design to perform multi-objective optimization to have minimum cost and stress and provided Pareto solution. 
\citet{BobbySpenceKareem2016J,SuksuwanSpence2018J} developed data driven simulation-based framework for topology optimization of tall buildings under wind and seismic environments. 
Some attempts have been made to utilize control strategy system during seismic events to optimize the dynamic behaviour of an integrated cable-stayed bridge \cite{FerreiraSimoes2011J}. 
\citet{BernardiniSpenceWeiKareem2015J} present a multi-objective aerodynamic shape optimization of tall buildings to reduce the lift and the drag coefficients by using an evolutionary algorithms and CFD simulations. 
\citet{CabanWhitemanPhillipsMastersDavisBridge2020J} presented automated cyber-physical system framework design and optimization for tall buildings by adjusting the dynamic properties of the structure. 
\citet{MartinsSimoesaNegraoa2020J} provided a comprehensive detailed review on optimization techniques to cable-stayed bridges under different dynamic load effects. 

\citet{SkinnerBehtash2018J} gave an overview of the dominant optimisation approaches for aerodynamic shape optimisation and classified into 6 different optimisation algorithm approaches for non-specialists. 
Optimization methods have been used for buildings to reduce wind loads by carrying out probabilistic performance-based assessment \cite{SpenceKareem2014J} and system-level reliability-based optimization \cite{SuksuwanSpence2019J}. 
A multi-objective optimization algorithm has been used to obtain Pareto optimal solutions for shape sculpting of tall building cross-section using CFD simulations \cite{DingKareem2018J,DingKareemWan2019J}. 
\citet{ElshaerBitsuamlakDamatty2017J} showed building corner aerodynamic optimization procedure by utilizing large eddy simulation (LES) and an artificial neural network (ANN) based surrogate model for drag and lift minimization for along- and across-wind responses. 
\citet{HorvatBrunoKhrisRaffaele2020J} used CFD simulations to maximize the aerodynamic performances of wind shields against windblown sand alongside line-like infrastructures. 
\citet{MaDuan2013J} have shown optimized wind screens for high-speed railways to protect the running trains from strong winds. 
\citet{RakocevicPopovic2018J} computed amplitudes of oscillation due to vortex shedding for stadium lighting pole to determine fatigue strength according to Eurocode based calculation methods. 

The optimization strategies have started to get some attention to be used in bridge aerodynamics in the last few years. 
\citet{WangDragomirescu2016J} tested a new type of multi-box bridge deck in the wind tunnel for static force coefficients, flutter derivatives and critical flutter wind speed. \citet{NietoHernandezJurado2009J,NietoMontoyaHernandezKusanoCasteleiroAlvarezJuradoFontan2020J} performed a parametric investigation on the slotted cross sections to apply the analytical optimum design problem formulation to optimize the cross section. 
Multi-objective optimization algorithm have been presented in  \cite{JaouadiAbbasMorgenthalLahmer2020J} to obtain Pareto optimal solutions for a bridge cross section by using CFD simulations to minimize the aerodynamic forces. 
\cite{TangLiWangTao2017J} used an optimization schemes for flutter performance of a suspension bridge with truss girder with horizontal and vertical aerodynamic countermeasures for the improvement of the bridge flutter stability. 
\cite{WangXiongWangGuoBaiLi2020J} performed numerical analysis for the assessment and design optimization method considering the flutter stability. 
\citet{KusanoBaldomirJuradoHernandez2014J,KusanoBaldomirJuradoHernandez2015J,KusanoMontoyaBaldomirNietoJuradoHernandez2020J}
utilized different reliability based design optimization methods for long-span bridges with a probabilistic flutter constraint to minimize the bridge girder weight. 
Some parametric studies have been made using CFD simulations and response surfaces \cite{MontoyaNietoHernandezKusanoAlvarezJurado2017J,MontoyaHernandezNietoKareem2020J} and were later used to optimize the deck cross section of bridges considering flutter \cite{MontoyaHernandezNieto2018J}. 
However, these studies only consider quasi-steady formulation for determining aerodynamic derivatives. 
A recent study \cite{MontoyaNietoHernandezFontanJuradoKareem2021J} provides aero-structural optimization of a bridge with short gap twin-box decks considering flutter and buffeting. 
However, it only uses Quasi-steady formulation for the aeroelastic analysis.

The aforementioned studies do not consider all the relevant phenomena necessary in the design of the long-span bridges. 
This paper provides a framework to perform aerodynamic shape optimization using CFD simulations and response surface techniques. 
The key element here is the use of response surface strategies to limit the number of CFD simulations to make the process more efficient. 
This offers to include all the essential global aeroelastic phenomena used in the design of the long-span bridges. 
Several target quantities are considered and checked in the optimization process as constraints which are required to be satisfied. 
This covers a wide range of phenomena and offers to simultaneously include even more phenomena based on the situation. 
The strategy is applied to a suspension bridge to optimize its deck shape where the CFD simulations have been used to determine the aerodynamic parameters and then semi-analytical approaches have been adopted to perform buffeting, flutter and Vortex-induced Vibrations (VIV) analyses. 
Here, the linear unsteady model has been utilized which not only takes into account the static wind forces but also the aerodynamic admittance and motion induced forces. 
To the best of authors knowledge, this has not been done before ans it covers the most relevant aerodynamic phenomena for the design of long-span bridges. 

\section{Description of the optimization framework}
\label{Section:DescriptionFramework}

\begin{figure*}[htbp]
	\centering
	\includegraphics[scale=0.9,trim=45mm 55mm 45mm 0mm,clip]{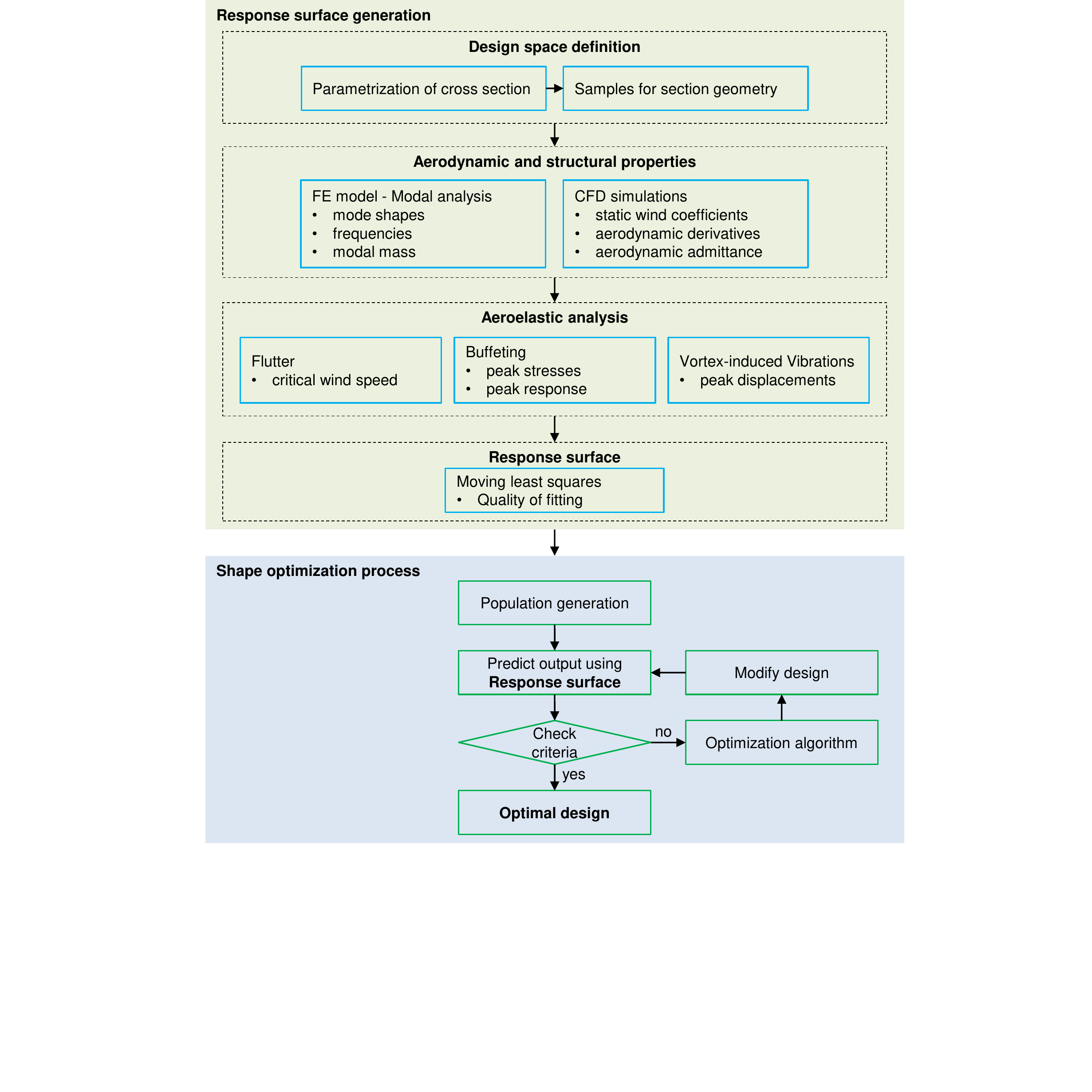}
	\caption{Framework for aerodynamic shape optimization.}
	\label{Figure:FlowChartFramework}
\end{figure*}

In the classical design process, the trial deck geometry is used to obtain the aerodynamic parameters from wind tunnel tests which are then utilized in the subsequent aerodynamic analyses. 
The structure is designed for buffeting forces and checked for the VIV response and the flutter onset. 
In case the design requirements are not achieved, the process is repeated with a modified cross section geometry. 
These shape modifications would require another cycle of wind tunnel testing hence increasing the design cost. 
However, by utilizing the advantages of CFD methods and semi-analytical approaches presented here, this process can be made efficient and cost effective. 

The optimization in the structural design context is an iterative process of modifying the shape of the structure with the aim to reduce a certain quantity of interest (objective) by satisfying given requirements. 
The target could be to reduce the weight of the structural members; however, the structure must be safe at the same time against the applied loads. 
This allows to select a design which is economical and effective to serve the purpose. 

The fundamental objective of the aerodynamic shape optimization for bridges is to modify the boundary of the deck cross section to achieve highest aerodynamic performance while satisfying the given constraints. 
The optimization process described here is based on the CFD simulations and numerical analyses. 
For this purpose, an efficient scheme is required to perform all the relevant analyses and to compute the required quantities of interest. 
This would also require an efficient data transfer between different analysis components. 
Since the CFD simulations are time intensive, the challenging aspect of this process is to develop an efficient but accurate numerical approach to perform the required aeroelastic analyses. 
This is achieved here using the advantages of the response surface approach. 

The proposed aerodynamic optimization strategy is depicted in Figure~\ref{Figure:FlowChartFramework}. 
The optimization process is summarized in the following.

The framework is divided into two main stages. 
The first stage of the process is to generate the response surface that will provide the target quantities for each sampled deck shape in the design domain. 
This step is performed before the main optimization algorithm and is mentioned as the 'Response surface generation' in the flowchart. 
In the second stage, the optimization of the cross section is performed using the response surface produced in the first stage. 
This step is mentioned as the `Shape optimization process'. 
The shape of the deck is controlled by the geometry parameters. 
In this step, each potential design candidate is checked for the relevant aeroelastic phenomena where the target quantities are computed through response surface from the first stage.

The deck geometry is parameterized which can be described with the help of design variables that are associated to the deck's shape. 
The samples for the cross section geometry using the design variables are generated in a specified design domain. 
For all sampled geometries, the subsequent analyses are performed to get aerodynamic and structural properties for the calibration of the response surfaces. 

A finite element model of the bridge is constructed and an eigenvalue analysis is performed for the structural properties of the bridge for each sample cross section. 
This provides mode shapes, natural frequencies and modal masses. 
Also, the aerodynamic properties are calculated for all the samples. 
The approach presented here utilizes CFD simulations to obtain aerodynamic parameters of the cross section geometries. 
These include static wind coefficients, aerodynamic derivatives and aerodynamic admittance functions. 

After obtaining the structural and aerodynamic properties, the analyses are performed for the relevant aeroelastic phenomena. 
The aeroelastic analyses comprise buffeting response, flutter and VIV. 
Further, a static analysis is performed to obtain stresses in the deck and the main cables under dead and live loads. 

The CFD simulations require a significant computational effort. 
To make the whole process efficient and more feasible, the response surface strategies have been utilized. 
The resulting quantities of interests are used to develop response surfaces. 
These response surfaces provide predictions for the buffeting forces, VIV response and flutter limits for the cross sections in the design domain. 

The response surfaces obtained are then used in the optimization process. 
The output of these response surfaces provide the target quantities which need to be checked against the limiting values. 
This is done for each candidate geometry in the optimization algorithm. 
This also requires several samples and iterations which commonly becomes computationally intensive.

It is possible to consider other constrains such as the seismic analysis requirements; however, it is beyond the scope of this paper. 
The central idea here is to provide a generally applicable framework to perform the aerodynamic shape optimization of the bridge deck cross section.

\section{Components of the framework}
\label{Section:ComponentsFramework}
This section elaborates the necessary components of the aerodynamic shape optimization framework used in this study. 
To keep the whole optimization framework efficient and achievable, a careful consideration has been given to perform the steps using advantages of parametrization and automation while ensuring careful data transfer between different process components without loosing useful information. 
This allows to set up the computational framework on a common desktop computer used nowadays.

\subsection{Structural modelling}
\label{Section:StructuralModelling}
The Finite Element (FE) type of idealisation provides a convenient and reliable analysis of the structural system. 
Further to that, dimensional reduction from a full FE analysis to modal superposition, for example, offers a superior efficiency. 
The displacement amplitudes are described using mode shapes which gives the number of degrees of freedom equal to the number of displacement components. 
This only requires few mode shapes to describe displacements with sufficient accuracy. 
The choice of mode selection depends on the type of the problem in hand. 
This reduces the size of the problem considerably. 
The natural frequencies and mode shapes are determined through modal analysis. 
Often the modes with the lowest frequencies are essential as they contribute most to the dynamic response. 
A numerical time integration scheme is used to solve the system under external forces in the time domain. 
The Newmark-Beta method \cite{CloughPenzien1993} is commonly used for this purpose.

\subsection{Aerodynamic parameters from CFD simulations}
\label{Section:AerodynamicParameters}
Aerodynamic parameters are required for the analyses of the bridge under different wind excitation phenomena. 
For this purpose, the aerodynamic characteristics are obtained through wind tunnel tests (WTT) or CFD simulations. 
The WTT can be expensive and time consuming. 
Despite the limitations, CFD approach provides an alternative to investigate wind effects on structures in relatively less time. 
Several parameters are required from 2D simulations on deck sections such as static wind coefficients, aerodynamic derivatives, Strouhal number and aerodynamic admittance functions. 
These aerodynamic parameters are obtained using the static and forced oscillation simulations on the bridge section. 
Some of the important parameters are defined as in the following. 

Bridge sections are bluff bodies where separation happens at even relatively low Reynolds number. 
The Reynolds number is the ratio between inertial forces to viscus forces defined as 
\begin{equation}
	\label{Equation:Reynolds_Number}
	Re=\frac{UB}{\nu},
\end{equation}
where $U$ is the wind speed, $B$ is the width of the section and $\nu$ is the kinematic viscosity. 

The integrated pressure on the cross section provides the aerodynamic forces in the global degrees of freedom. 
The non-dimensional aerodynamic forces are represented using static wind coefficients as: 
\begin{equation}
	C_{D}=\frac{F_D}{\frac{1}{2}U_{\infty}^{2}B}, \quad C_{L}=\frac{F_L}{\frac{1}{2}U_{\infty}^{2}B}, \quad C_{M}=\frac{F_M}{\frac{1}{2}U_{\infty}^{2}B^{2}},
\end{equation}
where $\rho$ is the air density, $C_{D}$, $C_{L}$ and $C_{M}$ are the coefficients of drag, lift and moment, respectively. Further,
$F_{D}$, $F_{L}$ and $F_{M}$ are the time averaged drag force, lift and moment, respectively. 

The Strouhal number is a non dimensional frequency of vortex shedding which allows to estimate the resonant wind speed for an oscillating structure under vortex-induced vibrations. 
The frequency of vortex shedding $f_v$ generated by the static bluff body is normalized as 
\begin{equation}
	\label{Equation:StrouhalNumber}
	St=\frac{f_vD}{U},
\end{equation}
where $St$ is the Strouhal number depending on the geometry of the body, $D$ is the across wind dimension of the body. 

The lateral force coefficient $c_{lat,0}$ of the deck cross section is used in the vortex-induced vibration analysis and is computed by obtaining the root mean square of the normalized lift force as 
\begin{equation}
	\label{Equation:LateralForceCoefficient}
	c_{lat,0}=\frac{\sigma({F_L})}{\frac{1}{2}U_{\infty}^{2}D},
\end{equation}
where $\sigma({F_L})$ is the standard deviation of the lift force. 

Aerodynamic derivatives have been used by Scanlan \cite{ScanlanTomko1971J} to describe the motion-induced aerodynamic forces. 
These are non dimensional coefficients which are functions of section geometry and reduced speed $v_r=U/Bf$ (see Section~\ref{Section:SemiAnalyticalModels}). 
These coefficients can be obtained by WTT on section models or through CFD simulations. 
Here the aerodynamic derivatives have been computed through forced vibration CFD simulations on the bridge section geometries. 
The detailed procedure to compute these parameters is presented in \citep{Abbas2016P}. 

Aerodynamic admittance $\chi$ represents the transfer function between the oncoming gust and the aerodynamic forces on the deck. 
Theoretical values of the aerodynamic admittance functions are available for a flat plate.  
For an arbitrary shape geometry, the aerodynamic admittance functions are obtained by performing CFD simulations with an oncoming deterministic gust on a static section. 
The concept implemented using numerical simulations has already shown encouraging results in \citet{Kavrakov2019P}. 
The same idea has been adopted in this framework for the section geometries to obtain the aerodynamic admittance functions. 

A flow solver \cite{Morgenthal2002} based on the Vortex Particle Method has been used to perform two dimensional CFD simulations to obtain the required aerodynamic parameters. 
As compared to classical Eulerian methods, it utilizes a grid-free Langragian formulation which makes it computationally efficient. 
A two dimensional slightly viscous incompressible flow is considered which is suitable for two dimensional bluff bodies. 
The Navier-Stokes equations are solved with the help of particles. 
The method can be applied to complex structural geometries. 
The boundary of the bluff cross section is defined as the input by dividing into small panels. 
The vorticity is discretised on the boundary of the body using the boundary element method. 
The pressure on the body surface is calculated based on the neighbouring velocity. 
Finally the integration of pressure provides forces in the global degrees of freedom. 
Considering the rigid section of the bridge, it offers to compute static wind coefficients, lateral force coefficient, aerodynamic derivatives and aerodynamic admittance functions. 

The solver has been tested and validated extensively for various benchmark cases and other bridge structures \citep{Abbas2016P,Morgenthal2002,KavrakovMorgenthal2018J,KavrakovMorgenthal2018J2,KavrakovMorgenthal2017J}. 
The accuracy of the vortex particle CFD simulations can be considered comparable with reference to the other methods such as Finite Volume Method. 
The comparison of the methods is not in the scope of this paper. 
The central idea here is to investigate the change in the aerodynamic behaviour of the section due to geometry variation and to quantify these effects to ultimately find the cross section with the best performance.

\subsection{Semi-analytical models for aeroelastic analysis}
\label{Section:SemiAnalyticalModels}
The four main actions to induce aerodynamic forces on bluff bodies are the static mean wind, vortex-induced excitation, the interaction between the wind and structural motion and the response due to wind fluctuations. 
These effects result in static mean wind forces, vortex-excited forces, motion-induced forces and buffeting forces, respectively \citep{GeXiang2008J2}. 

\begin{figure}[!htbp]
	\centering
	\includegraphics[scale=0.7,trim=0mm 0mm 0mm 0mm,clip]{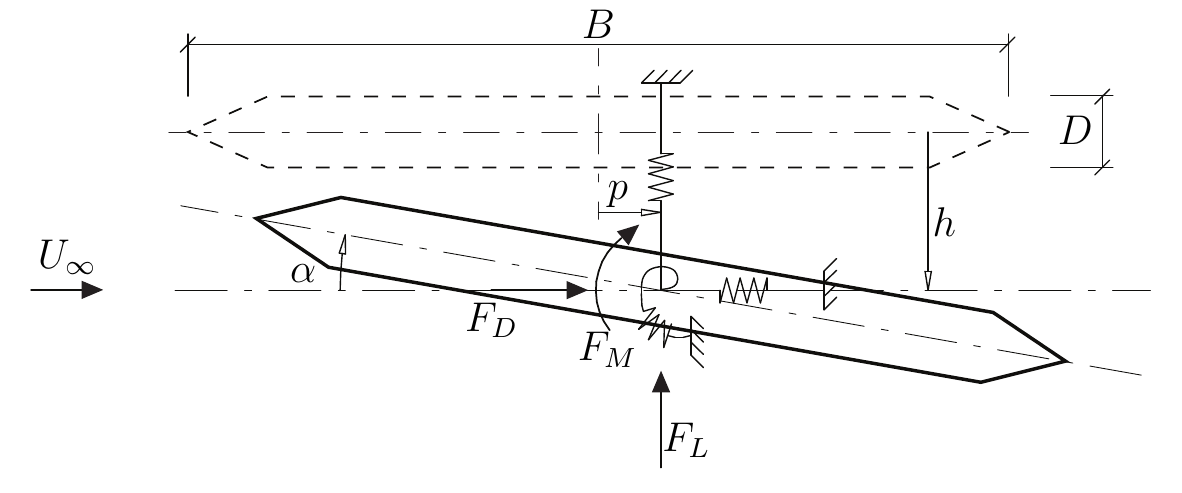}
	\caption{Definition of wind speed vectors, degrees of freedom and aerodynamic forces on the cross section.}
	\label{Figure:Figure_Schematic_Section_DOF}
\end{figure}

The idealisation of a two dimensional three degrees of freedom system is shown in Figure~\ref{Figure:Figure_Schematic_Section_DOF}. 
The equation of motions for the 3 degree of freedom system can be written as 
\begin{equation}
	\label{Equation:EquationOfMotion_Matrix_AE_B}
	\mathbf{M}\ddot{x}\left(t\right)+\mathbf{C}\dot{x}\left(t\right)+\mathbf{K}x\left(t\right)=\boldsymbol{F}_{ae}\left(t\right)+\boldsymbol{F}_{b}\left(t\right),
\end{equation}
where $\mathbf{M}$, $\mathbf{C}$, $\mathbf{K}$ are the mass, damping and stiffness matrices, respectively. 
$\boldsymbol{F}_{b}$ are the buffeting forces and $\boldsymbol{F}_{ae}$ are the motion-induced forces. 

\citet{Davenport1962J} included the unsteadiness in the buffeting forces for bridge sections by using aerodynamic admittance functions. 
The buffeting forces can be defined as 
\begin{subequations}
	\begin{align}
		F_{D}&=\frac{1}{2}\rho U_{\infty}^{2}B\left[2C_{D}\chi_{Du}\frac{u(t)}{U_{\text{\ensuremath{\infty}}}}+\left(C'_{D}-C_{L}\right)\chi_{Dw}\frac{w(t)}{U_{\text{\ensuremath{\infty}}}}\right],\\
		F_{L}&=\frac{1}{2}\rho U_{\infty}^{2}B\left[2C_{L}\chi_{Lu}\frac{u(t)}{U_{\text{\ensuremath{\infty}}}}+\left(C'_{L}+C_{D}\right)\chi_{Lw}\frac{w(t)}{U_{\text{\ensuremath{\infty}}}}\right],\\
		F_{M}&=\frac{1}{2}\rho U_{\infty}^{2}B^{2}\left[2C_{M}\chi_{Mu}\frac{u(t)}{U_{\text{\ensuremath{\infty}}}}+C'_{M}\chi_{Mw}\frac{w(t)}{U_{\text{\ensuremath{\infty}}}}\right],
	\end{align}
\end{subequations}
where $u(t)$, $w(t)$ are the fluctuating components of wind speed in the along wind and across wind direction, respectively. 
$\chi_{Du}$, $\chi_{Lu}$ and $\chi_{Mu}$ are the aerodynamic admittance coefficients in three directions. 

The eddies get separated from the body in a flow due to its motion. 
Motion-induced forces are produced as a result by countering the momentum contained in the eddies. 
The forcing depends on the history of the motion or the so-called fluid memory effect. 
Theodorsen investigated the flutter phenomenon for aircraft wings and gave motion-induced aerodynamic forces on a flat plate using circulation functions \cite{Theodorsen1935R}. 
The fluid-structure interaction for bluff sections is a complex phenomenon and analytical models can not describe motion-induced forces for these sections. 
This is due to the flow separation and reattachment which could occur at multiple points on the section boundary, and it contradicts the assumption for the shedding of vertices from a single point. 
\cite{ScanlanTomko1971J} introduced the motion-induced forces by using aerodynamic derivatives obtained from experimental approach. 
Scanlan's model has been extensively used in bridge aerodynamics for the solution of aeroelastic instability problems due to its applications to different types of bridge cross sections. 
The extended expression of Scanlan's model is represented as linear function of the motion considered in Theodorsen theory in terms of aerodynamic derivatives as 

\begin{align}
	\label{Equation:Scanlan_3DOF}
	F_{L}=&\frac{1}{2}\rho U_{\infty}^{2}B\biggl[KH_{1}^{*}\frac{\dot{h}}{U_{\infty}}+KH_{2}^{*}\frac{B\dot{\alpha}}{U_{\infty}}\\ \nonumber
	&+K^{2}H_{3}^{*}\alpha+K^{2}H_{4}^{*}\frac{h}{B}+KH_{5}^{*}\frac{\dot{p}}{U_{\infty}}+K^{2}H_{6}^{*}\frac{p}{B}\biggr],\label{Equation:Scanlan_Lift_3DOF}\\
	F_{M}=&\frac{1}{2}\rho U_{\infty}^{2}B^{2}\biggl[KA_{1}^{*}\frac{\dot{h}}{U_{\infty}}+KA_{2}^{*}\frac{B\dot{\alpha}}{U_{\infty}}\\ \nonumber
	&+K^{2}A_{3}^{*}\alpha+K^{2}A_{4}^{*}\frac{h}{B}+KA_{5}^{*}\frac{\dot{p}}{U_{\infty}}+KA_{6}^{*}\frac{p}{B}\biggr],\label{Equation:Scanlan_Moment_3DOF}\\
	F_{D}=&\frac{1}{2}\rho U_{\infty}^{2}B\biggl[KP_{1}^{*}\frac{\dot{p}}{U_{\infty}}+KP_{2}^{*}\frac{B\dot{\alpha}}{U_{\infty}}\\ \nonumber
	&+K^{2}P_{3}^{*}\alpha+K^{2}P_{4}^{*}\frac{p}{B}+KP_{5}^{*}\frac{\dot{h}}{U_{\infty}}+K^{2}P_{6}^{*}\frac{h}{B}\biggr],\label{Equation:Scanlan_Drag_3DOF}
\end{align}
where $F_D$ is the aerodynamic drag force on the section, $H_i^*$, $A_i^*$ and $P_i^*$ ($i=1,\ldots,6$) are the non-dimensional functions of $K$ known as aerodynamic or flutter derivatives which are associated with self-excited lift, moment and drag, respectively. 
$h$, $\alpha$ and $p$ are heave, pitch and horizontal displacement, respectively. 
$\dot{h}$, $\dot{\alpha}$ and $\dot{p}$ are the vertical, rotational and horizontal velocity, respectively. 
$K=B\omega/U$ is the reduced frequency and $\omega$ is the frequency of bridge oscillation under aerodynamic forcing. 
These aerodynamic derivatives for a bluff section are obtained by experimental wind tunnel tests or through the CFD simulations.

\subsection{Aeroelastic analyses}
\label{Section:AeroelasticAnalyses}

\subsubsection*{Buffeting response}
\label{Section:BuffetingResponse}
The bridge structure is subjected to the static mean wind forces and the dynamic wind forces due to fluctuations in the wind speed. 
Buffeting response is caused by fluctuating forces induced due to turbulence in the wind. 
It is considered an essential design criterion to determine the size of structural members at the design stage. 
To accurately predict the aeroelastic response of long-span bridges, motion-induced forces are also considered in addition to the buffeting forces. 
The buffeting analysis can be performed in the time domain or the frequency domain \citep{Xu2013B}. 
The time domain analysis has the benefit of capturing structural and aerodynamic nonlinearities as well as non stationary behaviour of the oncoming wind. 

The frequency dependent motion-induced forces presented in Eq.~\ref{Equation:Scanlan_3DOF} are formulated in the time-domain by indicial functions or convolution integrals. 
These indicial functions are obtained from WTT, CFD simulations or from aerodynamic derivatives. 
Alternatively, impulse function approach is also utilized. 
These indicial and impulse functions represent aerodynamic derivatives in the time domain and thus accounting for the fluid memory effects. 
They are obtained by utilizing rational approximation. 
The buffeting forces can be expressed in the form of convolution integrals with aerodynamic impulse functions and fluctuating wind speed components. 
The motion-induced forces are considered to be full-correlated in an element whereas span-wise correlation for buffeting forces is taken into account by joint acceptance function. 
The solution of the equation of motion for the whole system is obtained through Newmark-Beta time integration scheme \cite{Kavrakov2019P}. 

\subsubsection*{Flutter}
\label{Section:Flutter}
Flutter phenomenon is a coupling of aerodynamic forcing and structural dynamics. 
The analysis is commonly performed using complex eigenvalue solution. 
The motion-induced aerodynamic forces are functions of aerodynamic derivatives which depend on the frequency of oscillation. 
Therefore, the solution is performed in the frequency domain mainly due to the computational efficiency offered by the frequency domain. 
This is done by representing aeroelastic instability as an eigenvalue problem. 
The equation of motion is solved through eigenvalues. 
The response is assumed sinusoidal with a constant amplitude at the flutter boundary. 
By increasing wind speed $U_\infty$, the system is solved successively and becomes unstable when at least one eigenvalue has a positive real part indicating a divergent response. 
At this condition, the wind speed is considered to be the flutter limit $U_{cr}$. 
The implementation and further details are explained in \citep{Abbas2016P}. 
The sensitivity of the flutter limit to different input parameters can be found in \cite{AbbasProbabilistic2015J}. 

Commonly the streamlined cross sections exhibit classical flutter. 
However, a bluff or deeper cross section maybe prone to torsional flutter. 
The stability of such a cross section is associated to the aerodynamic derivative $A_{2}^*$. 
When $A_{2}^*$ becomes positive, the total damping will become negative that will cause the system to oscillate with divergent amplitudes. 
The instability condition of the SDOF torsional flutter \citep{Abbas2016P} can be expressed as: 
\begin{equation}
	\label{Equation:TorsaionalFLutter}
	A_{2}^*=\frac{4m_{\alpha}\xi_{\alpha}}{\rho B^{4}},
\end{equation}
where $m_{\alpha}$ is the torsional mass, and $\xi_{\alpha}$ is the damping ratio to critical for the torsional mode. 

\subsubsection*{Vortex-induced vibrations}
\label{Section:VortexInducedVibrations}
A bluff body in a fluid flow creates vortices behind the body as a result of separation of the boundary layer. 
The shedding causes unsteady forces in the body perpendicular to the direction of the flow. 
The pattern of vortex shedding depends on the velocity of flow and the geometry of the body. 
If the shedding frequency is close to the natural frequency, the oscillations of the body take control of the vortex shedding and resonance may occur. This control of the situation by the mechanical forces is called Lock-in. 
The vibration become considerably large for a certain band of wind speeds. 
The wind speed at which it starts to happen is known as resonant wind speed $U_{res}$. 
\begin{equation}
	U_{\mathrm{res}, \mathrm{i}}=\frac{D \cdot f_{\mathrm{ni}, \mathrm{y}}}{\mathrm{St}}.
\end{equation}
This leads to serviceability or ultimate limit state due to fatigue over a longer period of time. 
Eurocode \cite{Eurocode2004} provides an approach based on the Ruscheweyh model \citep{Sockel1994B} which allows to compute the peak displacements due to VIV response as 
\begin{equation}
	\label{Equation:EurocodeVIV}
	\frac{h_{\max }}{D}=\frac{1}{S t^{2}} \cdot \frac{1}{S c} \cdot K \cdot K_{w} \cdot c_{\text {lat }}
\end{equation}
where $St$ is the Strouhal number (see Eq.~\ref{Equation:StrouhalNumber}). 
$Sc$ is the Scruton number which is defined as 
\begin{equation}
	Sc=\frac{2\cdot\delta_{s}\cdot m_{i,e}}{\rho\cdot D^{2}},
\end{equation}
where $K_W$ and $K$ are the effective correlation length factor and the mode shape factor, respectively. 
These are defined as 
\begin{align}
	K_{\mathrm{w}}&=\frac{\sum\limits_{j=1}^{n} \int\left|\Phi_{i, y}(s)\right| \mathrm{d} s}{\sum\limits_{j=1}^{m} \int_{l_{\mathrm{j}}}\left|\Phi_{i, y}(s)\right| \mathrm{d} s} \leq 0.6,\\
	K&=\frac{\sum\limits_{j=1}^{m} \int_{\ell_{j}}\left|\Phi_{i, y}(s)\right| d s}{4 \cdot \pi \cdot \sum\limits_{j=1}^{m} \int_{l_{j}} \Phi_{i, y}^{2}(s) d s}
\end{align}
where $c_{lat}$ is the lateral force coefficient (see Eq.~\ref{Equation:LateralForceCoefficient}), 
$\Phi_{i, y}$ is the mode shape $i$, 
$l_j$ is the correlation length, 
$j$ is the length of the structure between two nodes, 
$n$ is the number of regions where vortex excitation occurs at the same time, 
$m$ is the number of antinodes of the vibrating structure in the considered mode shape $\Phi_{i, y}$, 
$s$ is the coordinates of the structure along length. 

The mechanism of vortex shedding is not uniformly distributed along the bridge deck. 
The excitation forces are highly correlated at the antinodes of the mode shape. 
The effective correlation length factor $K_W$ allows to take into account the aeroelastic forces. 
For further details about the method, the reader is referred to \cite{Sockel1994B}.

\subsection{Response surface}
\label{Section:ResponseSurface}
The CFD simulations require a significant computational time. 
Response surface strategies may be used in this case to make the process more efficient. 
Surrogate modelling is viewed as a nonlinear inverse problem where the target is to achieve a continuous function of model response based on a limited data. 
This surrogate model can provide predictions for the missing parameter space for which the CFD simulations have not been performed which is a significant advantage. 
A CFD simulation may require a few days to complete whereas running the response surface would need only a fraction of this time which greatly reduces the computational cost without modifying the existing numerical solver for the simulations. 
The choice of the response surface model is essential to obtain accurate predictions. 
A reasonable quality of the surrogate model can be achieved if a sufficient number of numerical simulations is included with suitable combinations of input parameters. 

Polynomial regression utilizes polynomial basis function to approximate the model response. 
The approach has limitations in case of highly nonlinear model to capture the response efficiently. 
The Moving Lest-Squares approach has been utilized here that has the ability to also capture localised regions of the model response. 
This is done by introducing radial weighting functions which are associated to the location of the data point being evaluated. This is in contrast to polynomial response surface approach where equal weights are considered for all data points. 
It may require a greater computational effort but provides superior behaviour \cite{LancasterSalkauskas1981J}. 

The sampling plan is an essential step for an accurate response surface generation. 
The number of required samples is related to the number and range of parameters as well as nonlinearity of the mode behaviour. 
The validation of the response surface is therefore important before using the response surface for the subsequent analyses. 
One viable and effective option is to implement cross-validation techniques. 

\subsection{Optimization strategy}
\label{Section:OptimizationStrategy}
The target of optimization process is to minimize a given objective function or a property while satisfying the specified constraints. 
Commonly the cost is the target for structural optimization which is in the form of weight of the structure that needs to be reduced and the constraints are the limiting aeroelastic phenomena. At the same time it is made sure that the dimensions are within the allowable limits.

The objective function ${f}$ for the aerodynamic optimization is represented as 
\begin{equation}
	\min f\left(\bf{x}\right)=\min f\left(d_x,b_x\right)=m\left(d_x,b_x\right)\times L_{span}
\end{equation}
where ${f}$ is the objective function required to be minimized and $\bf{x}$ is a vector containing design variables. 
$\bf{x}$ can be geometry parameters such as $d_x,b_x$ related to depth $D$ and width $B$, respectively as explained in Section~\ref{Section:DeckShapeParametrization}. 
The optimization algorithm provides values of these parameters for which the objective function is minimum. 
The design variables are often constrained such as 
\begin{equation}
	\bf{x_{min}}\leq\bf{x}\leq\bf{x_{max}}
\end{equation}
where $\bf{x_{min}}$ and $\bf{x_{max}}$ represent the minimum and maximum limits of the design variables. 
Constraints are introduced in the process to contain the optimization problem by producing design within specified limits. 
The constraints considered for the problem can be represented by the following: 
\begin{subequations}
	\begin{align}
		g_{i}^{P}\left(\bf{x}\right)&=P_{i}-P_{max}\leq0, \quad i=1,\ldots,s \quad \textrm{(Case I)}\\
		g_{i}^{P}\left(\bf{x}\right)&=P_{min}-P_{i}\leq0, \quad i=1,\ldots,s \quad \textrm{(Case II)}
	\end{align}
	\label{Equation:GeneralConstraints}
\end{subequations}
where $g_{i}^{P}$ represent the imposed constraint, $P_{i}$ are the computed quantities of interest, $i$ is the number of scenarios or particular control point, $s$ are the number of control points or considered elements or scenarios (see Table~\ref{Table:Constraints}), $P_{max}$ and $P_{min}$ are the limiting values of the quantities, respectively. 
Case I and II represent the constraints for the quantities which need to be smaller or larger than the allowable limits, respectively. 
This simplification of constraints is a useful step to evaluate objective functions efficiently. 

There are several methods available to treat the optimization problem. 
They can broadly be classified into gradient-based and stochastic approaches. 
The former consists of computing slopes of the objective function with respect to the input variables. 
The process is repeated iteratively to achieve convergence. 
The latter comprises approaches which do not require to compute slopes such as genetic algorithms \cite{Simon2013B} and swarm algorithms \cite{WangTanLiu2018J}. 
Here in this study, the stochastic optimization approaches have been used with constraints. 
Particle swarm optimization has been adopted for this purpose. 
This model is inspired from the behaviour of bird flocking in nature. 
\citet{WangTanLiu2018J} provide an overview and background of the PSO. 

\citet{EberhartKennedy1995C} developed an evolutionary optimization by considering each participant as a particle without mass and volume with just position and velocity hence `particle swarm optimization'. 
Each particle is considered as a potential solution for optimization. 
It can memorize the best position and velocity of the swarm and itself. 
In every iteration, the velocity information is combined to calculate the new position for each particle. 
These particles continue to change their positions in the N-dimensional domain until an optimal position is reached. 

\begin{figure}[!htbp]
	\centering
	\includegraphics[scale=0.75,trim=18mm 0mm 0mm 0mm,clip]{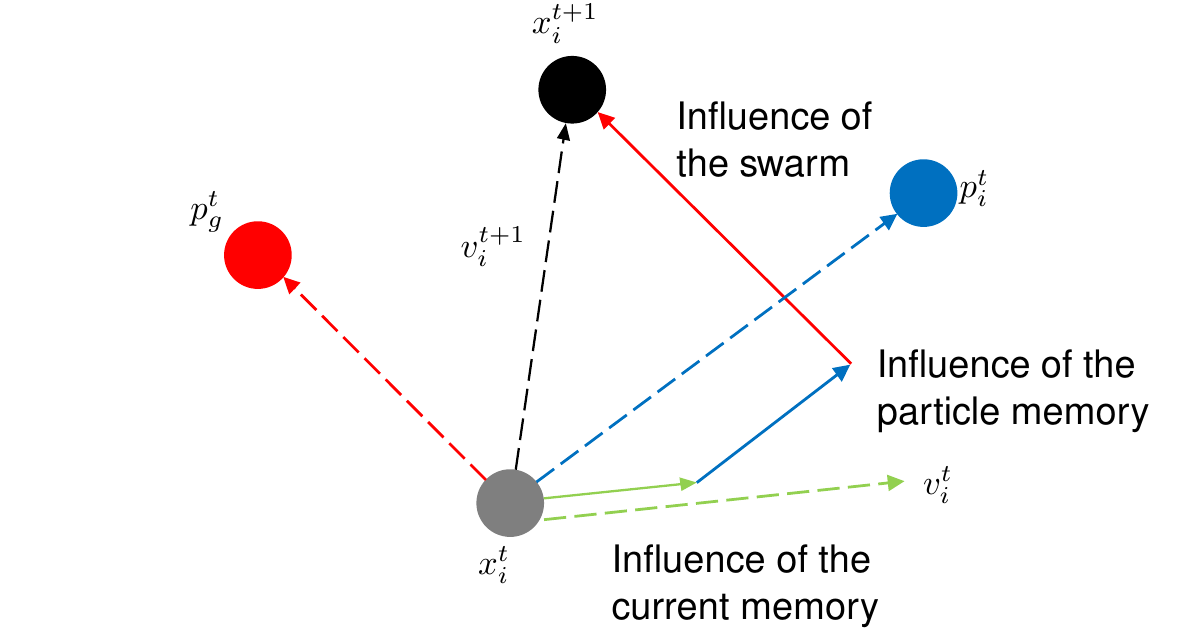}
	\caption{Iteration scheme of the particles in the PSO (after \cite{WangTanLiu2018J}).}
	\label{Figure:ParticleSwarmSchematic}
\end{figure}

A schematic of the PSO iteration step is presented in Figure~\ref{Figure:ParticleSwarmSchematic}. 
Each particle is represented as a point in the Cartesian coordinate system which is assigned in the beginning with a random initial position ($x^t_i$) and an initial velocity ($v^t_i$). 
Cost is calculated which represents a difference to the optimum solution at these positions. 
Each point can memorize its best position ($p^t_i$). 
Also it is considered that the swarm can communicate among each other and inform about the best position ($p^t_g$) of the total swarm. 
Based on the location of each particle from personal best ($p^t_i$) and global best ($p^t_g$), the velocity is scaled randomly. 
A weighting factor on the personal best ($2r_2$) and the global best ($2r_3$) velocities decide how fast the particles will move towards the personal best ($p^t_i$) or the global best ($p^t_g$). 
The new velocity ($v_{i}^{t+1}$) is computed as 
\begin{equation}
	v_{i}^{t+1}=2 r_{1} v_{i}^{t}+2 r_{2}\left(p_{i}^{t}-x_{i}^{t}\right)+2 r_{3}\left(p_g^{t}-x_{i}^{t}\right).
\end{equation}
A new position ($x^{t+1}_i$) is computed and the particle is moved to the new location using 
\begin{equation}
	x_{i}^{t+1}=x_{i}^{t}+v_{i}^{t+1}.
\end{equation}

\section{Application of aerodynamic optimization framework}
\label{Section:ApplicationFramework}
The workflow is developed using MATLAB which allows to generate samples, create geometries, generate finite element models, run CFD simulations and aeroelastic analyses using semi-analytical models, and finally to develop response surfaces (training and validation). 
The optimization algorithm is also implemented in the same environment. 

\subsection{Reference structure}
\label{Section:ReferenceStructure}
\begin{figure*}[!htbp]
	\centering
	\includegraphics[scale=0.9,trim=0mm 0mm 0mm 0mm,clip]{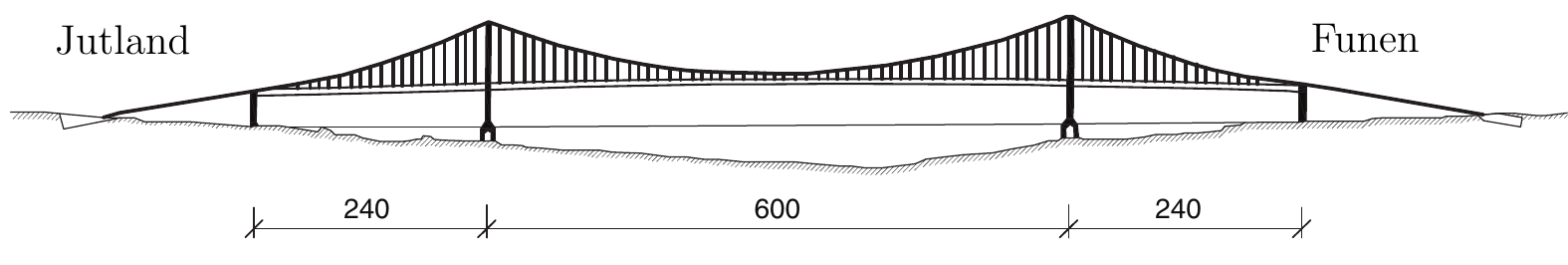}
	\caption{Reference suspension bridge (units: [m]).}
	\label{Figure:Bridge_Elevation}
\end{figure*}

The Lilleb\ae lt Suspension Bridge, Denmark has been used as a reference object within the study. 
The elevation of the bridge is presented in Figure~\ref{Figure:Bridge_Elevation}. 
The Lilleb\ae lt Suspension Bridge is the least studied bridge in the bridge aerodynamics. It is smaller but important predecessors to the Greatbelt Bridge being one of the earliest example of streamlined sections in Europe \cite{Scanlan1990J}. 
This was another motivation to explore the structure for the aerodynamic performance utilising the modern analysis tools. 

The Lilleb\ae lt Suspension Bridge consists of a steel box girder deck with 1080~m total suspended span. 
A 1:200 scaled bridge model was developed during construction to get the dynamic propertied of the bridge. 
The details can be found in the design reports \cite{Publication_IIIR,Publication_XIR}. 
The deck is discontinues at the tower location unlike its successor the Greatbelt Suspension Bridge with a continues steel deck. 
Main dimensions and some selected properties are presented in Table~\ref{Table:Bridge_properties}. 
Figure~\ref{Figure:ReferenceSection} shows the simplified cross section of the bridge considered in this study. 
For simplicity, railings and other attachments on the sections are not considered. 
This was done to reduce the computational effort for the CFD simulations. 

\begin{table}[!htbp]
	\caption{Properties of the reference bridge.}
	\label{Table:Bridge_properties}
	\centering
	\begin{tabularx}{0.5\textwidth}{L{0.8}R{0.2}}
		\hline
		Total suspended span $L$ [m]				& 1080\\
		Main span length [m] 						& 600\\
		Side span length [m]						& 240\\
		Tower height [m] 							& 112\\
		Deck navigation clearance [m]				& 44\\				
		Main cable sag [m]							& 67\\	
		Main cable spacing [m]						& 28.1\\
		Suspender spacing [m]						& 12\\
		Main cable steel area [mm$^2$] 				& 159500\\
		Suspender steel area [mm$^2$]				& 2$\times$1840\\
		Deck width $B$ [m]						    & 33.3\\
		Deck depth $D$ [m]						    & 3\\
		Translational mass $m_h$, $m_p$ [t/m]		& 11.7\\
		Mass moment of inertia $m_\alpha$ [tm$^2$/m]& 1063.9\\
		Damping ratio to critical $\xi$ [$\%$]		& 0.5\\
		\hline
	\end{tabularx}
\end{table}

\begin{figure*}[!htbp]
	\centering
	\includegraphics[scale=0.9,trim=15mm 0mm 15mm 0mm,clip]{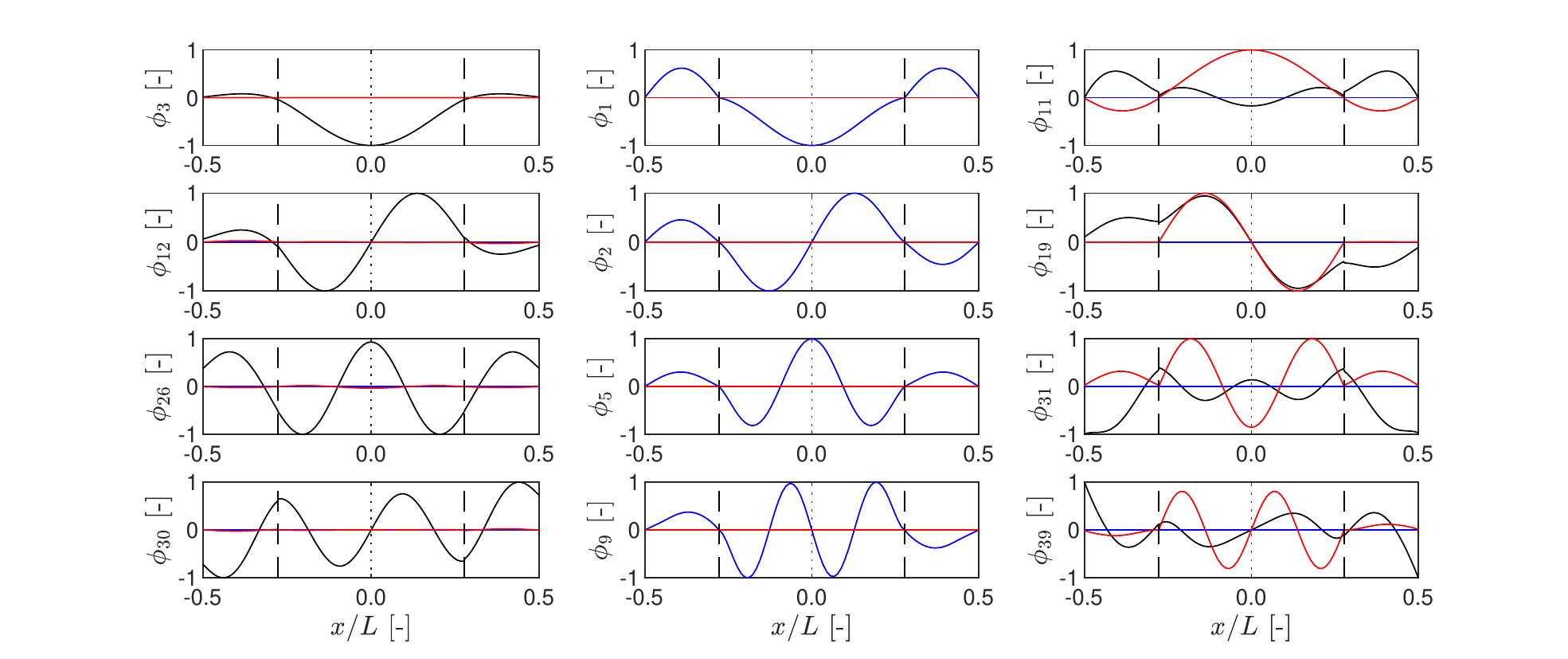}
	\caption{Sample normalized mode shapes of the bridge deck considering the reference cross section: (left) lateral mode shapes, $f_{3}=0.219$Hz, $f_{12}=0.577$Hz, $f_{26}=0.995$Hz, $f_{30}=1.159$Hz, (middle) vertical, $f_{1}=0.156$Hz, $f_{2}=0.197$Hz, $f_{5}=0.300$Hz, $f_{9}=0.468$Hz, (right) torsional, $f_{11}=0.500$Hz, $f_{19}=0.776$Hz, $f_{31}=1.166$Hz, $f_{39}=1.532$Hz, (dash lines indicate tower locations, deck centre at $x/L=0.0$).}
	\label{Figure:Step5_ModeShapes}
\end{figure*}

A finite element model of the Lilleb\ae lt Suspension Bridge is developed to obtain the structural properties. 
The model is based on the information provided in \cite{Publication_IIIR,Publication_XIR}. The model is parametrized so that the cross section details can be changed easily for the subsequent analyses. 
The concrete pylons and the main steel deck have been modelled with beam elements. 
The main cable and suspenders have been modelled with cable elements. 
Springs are used at the deck-tower connection locations and at the base of the tower. 
The model is first calibrated by varying the cable pre-tension and spring stiffness at the joints to target the frequencies provided in \cite{Publication_XIR}. 
The frequencies for the first bending and the first torsional modes have been achieved reasonably well. 
The mode shapes and the corresponding frequencies are depicted in Figure \ref{Figure:Step5_ModeShapes}. 


\subsection{Deck shape parametrization and design space definition}
\label{Section:DeckShapeParametrization}
\begin{figure*}[!htbp]
	\centering
	\includegraphics[scale=0.9,trim=15mm 0mm 0mm 0mm,clip]{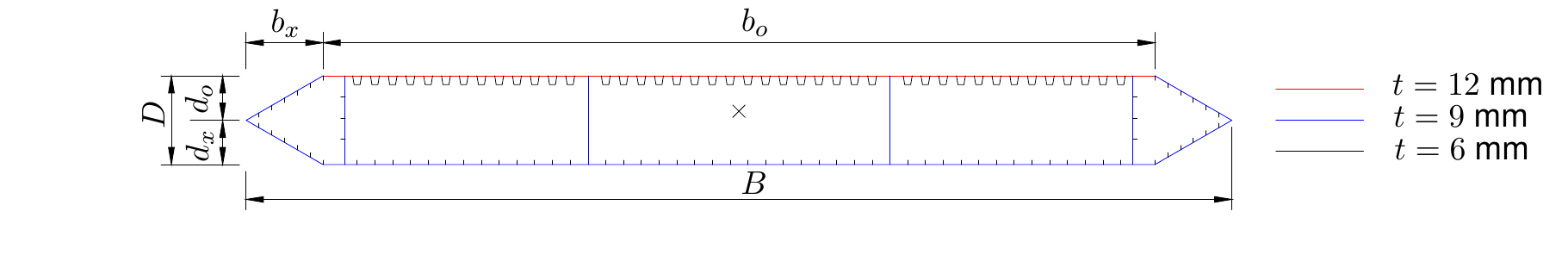}
	\caption{Parametrization of the deck geometry (for the reference section $b_x=2.6$m, $d_x=1.5$m).}
	\label{Figure:ReferenceSection}
\end{figure*}

\begin{figure*}[!htbp]
	\centering
	\includegraphics[scale=0.9,trim=5mm 20mm 5mm 25mm,clip]{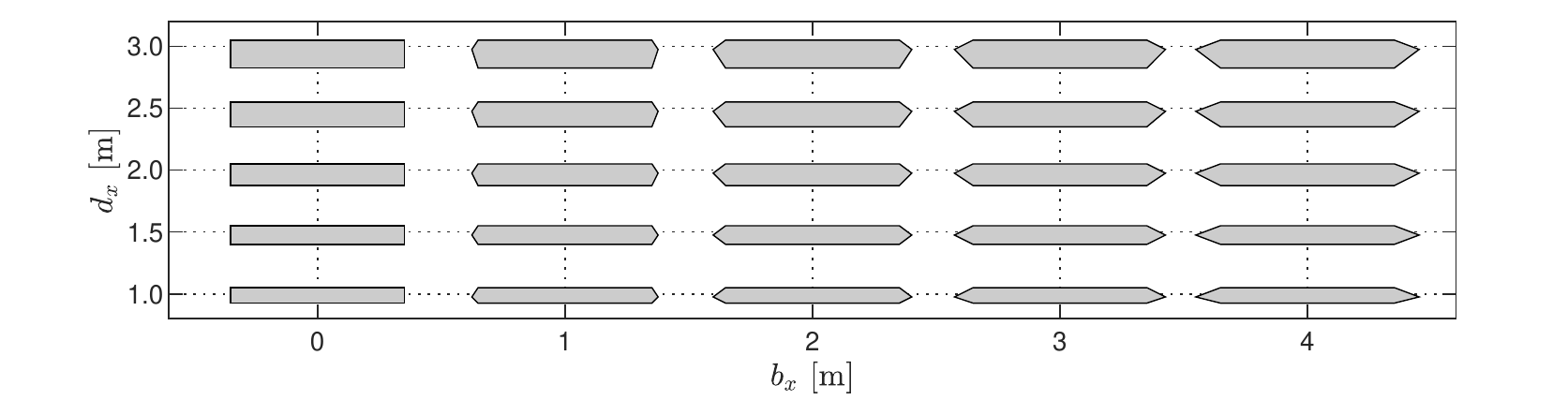}
	\caption{Design space: sampled deck shapes used in the response surface generation (see Figure~\ref{Figure:ReferenceSection}).}
	\label{Figure:Step2_GeometryVariation}
\end{figure*}

The idea of the optimization process followed here is to obtain a cross section with least weight and that can satisfy some constraints. 
A reference section is considered in the analyses just for comparison. 
The cross section geometry as presented in Figure~\ref{Figure:ReferenceSection} is parametrized using control variables which can produce a wide range of cross sectional shapes. 
The cross section consists of several plate segments with different plate thicknesses. 
The inner details with stiffening elements are modelled to obtain a more realistic stiffness distribution in the cross section. 
The central top part of the section (road width $b_o$) and its plate thickness are kept constant. 
Also, the distance from top to the fairing edge $d_o$ is considered as fixed. 
The fairing angle and the depth can be changed by modifying the values of $d_x$ and $b_x$. 
By only changing these two parameters, the section geometry is modified significantly covering a wide range of deck shapes as shown in Figure~\ref{Figure:Step2_GeometryVariation}. 

A sampling plan in the input parameter space is required to build the response surface, where a regular arrangement of the samples within equidistant points is chosen.
It is apparent that the aerodynamic performance of the rectangular section is inferior as compared to the more streamlined cross section; however, it was kept in the sample set to have a comparison. 
The generated cross section geometries have been then used to compute the required parameters as summarized in Table~\ref{Table:Mechanical_Properties}.

\begin{table*}[htbp]
	\centering
	\small
	\caption{Mechanical properties of the sampled bridge cross sections (with plate thickness on Figure~\ref{Figure:ReferenceSection}). ($d_c$: centroid along y-axis, $A$: area of steel, $I_y$, $I_z$: second moment of area about y and z-axes, $I_t$: polar moment of inertia, $m_h$, $m_p$: translational mass in the vertical and lateral directions, $m_\alpha$: torsional mass).}
	\label{Table:Mechanical_Properties}
	\begin{tabularx}{\textwidth}{L{0.15}R{0.1}R{0.1}R{0.1}R{0.1}R{0.1}R{0.1}R{0.1}R{0.1}R{0.15}R{0.15}R{0.15}R{0.15}}
		\hline
		Sample & $d_x$ & $b_x$ & $B$ & $D$ & $B/D$ & $d_c$ & $A$ & $I_y$ & $I_z$ & $I_t$ & $m_h$, $m_p$ & $m_\alpha$\\
		\ [-] & [m] & [m] & [m] & [m] & [-] & [m] & [m$^2$] & [m$^4$] & [m$^4$] & [m$^4$] & [t/m] & [ktm/m]\\
		\hline
		Ref. & 1.5 & 2.6 & 33.3 & 3.0 & 11.1 & 1.19 & 1.11 & 1.86 & 99.3 & 4.99 & 11.68 & 1.064\\
		\hline
		1	&	1.0	&	0.0	&	28.1	&	2.5	&	11.2	&	0.97	&	1.01	&	1.22	&	76.7	&	3.28	&	10.62	&	0.820	\\
		2	&	1.0	&	1.0	&	30.1	&	2.5	&	12.0	&	0.98	&	1.02	&	1.23	&	80.2	&	3.43	&	10.75	&	0.856	\\
		3	&	1.0	&	2.0	&	32.1	&	2.5	&	12.8	&	0.99	&	1.06	&	1.26	&	88.9	&	3.48	&	11.12	&	0.949	\\
		4	&	1.0	&	3.0	&	34.1	&	2.5	&	13.6	&	1.01	&	1.09	&	1.28	&	99.6	&	3.50	&	11.51	&	1.061	\\
		5	&	1.0	&	4.0	&	36.1	&	2.5	&	14.4	&	1.02	&	1.14	&	1.31	&	112.4	&	3.52	&	11.94	&	1.196	\\
		6	&	1.5	&	0.0	&	28.1	&	3.0	&	9.4	&	1.17	&	1.05	&	1.81	&	81.6	&	4.64	&	11.00	&	0.877	\\
		7	&	1.5	&	1.0	&	30.1	&	3.0	&	10.0	&	1.18	&	1.06	&	1.82	&	84.8	&	4.87	&	11.12	&	0.911	\\
		8	&	1.5	&	2.0	&	32.1	&	3.0	&	10.7	&	1.19	&	1.09	&	1.84	&	93.2	&	4.96	&	11.46	&	1.000	\\
		9	&	1.5	&	3.0	&	34.1	&	3.0	&	11.4	&	1.20	&	1.12	&	1.87	&	103.2	&	5.01	&	11.82	&	1.105	\\
		10	&	1.5	&	4.0	&	36.1	&	3.0	&	12.0	&	1.21	&	1.16	&	1.91	&	115.8	&	5.04	&	12.24	&	1.239	\\
		11	&	2.0	&	0.0	&	28.1	&	3.5	&	8.0	&	1.38	&	1.08	&	2.52	&	86.2	&	6.20	&	11.37	&	0.933	\\
		12	&	2.0	&	1.0	&	30.1	&	3.5	&	8.6	&	1.38	&	1.09	&	2.53	&	89.2	&	6.52	&	11.47	&	0.965	\\
		13	&	2.0	&	2.0	&	32.1	&	3.5	&	9.2	&	1.38	&	1.12	&	2.56	&	97.0	&	6.68	&	11.77	&	1.047	\\
		14	&	2.0	&	3.0	&	34.1	&	3.5	&	9.7	&	1.39	&	1.15	&	2.59	&	107.1	&	6.76	&	12.14	&	1.154	\\
		15	&	2.0	&	4.0	&	36.1	&	3.5	&	10.3	&	1.39	&	1.19	&	2.63	&	119.2	&	6.82	&	12.53	&	1.282	\\
		16	&	2.5	&	0.0	&	28.1	&	4.0	&	7.0	&	1.58	&	1.12	&	3.36	&	91.0	&	7.96	&	11.75	&	0.993	\\
		17	&	2.5	&	1.0	&	30.1	&	4.0	&	7.5	&	1.58	&	1.13	&	3.37	&	94.1	&	8.39	&	11.85	&	1.026	\\
		18	&	2.5	&	2.0	&	32.1	&	4.0	&	8.0	&	1.58	&	1.15	&	3.40	&	101.6	&	8.62	&	12.13	&	1.105	\\
		19	&	2.5	&	3.0	&	34.1	&	4.0	&	8.5	&	1.58	&	1.18	&	3.45	&	111.2	&	8.75	&	12.46	&	1.206	\\
		20	&	2.5	&	4.0	&	36.1	&	4.0	&	9.0	&	1.59	&	1.22	&	3.50	&	123.6	&	8.84	&	12.87	&	1.337	\\
		21	&	3.0	&	0.0	&	28.1	&	4.5	&	6.2	&	1.79	&	1.15	&	4.34	&	95.9	&	9.90	&	12.14	&	1.055	\\
		22	&	3.0	&	1.0	&	30.1	&	4.5	&	6.7	&	1.79	&	1.16	&	4.35	&	99.1	&	10.44	&	12.23	&	1.088	\\
		23	&	3.0	&	2.0	&	32.1	&	4.5	&	7.1	&	1.79	&	1.19	&	4.39	&	106.5	&	10.76	&	12.50	&	1.166	\\
		24	&	3.0	&	3.0	&	34.1	&	4.5	&	7.6	&	1.79	&	1.22	&	4.44	&	115.9	&	10.95	&	12.82	&	1.266	\\
		25	&	3.0	&	4.0	&	36.1	&	4.5	&	8.0	&	1.79	&	1.26	&	4.51	&	128.1	&	11.09	&	13.21	&	1.395	\\
		\hline																									
		Max.	&	3.0	&	4.0	&	36.1	&	4.5	&	14.4	&	1.79	&	1.26	&	4.51	&	128.1	&	11.09	&	13.21	&	1.395	\\
		Min.	&	1.0	&	0.0	&	28.1	&	2.5	&	6.2	&	0.97	&	1.01	&	1.22	&	76.7	&	3.28	&	10.62	&	0.820	\\
		\hline
	\end{tabularx}
\end{table*}

\subsection{Structural characteristics and aerodynamic properties from CFD Simulations}
\label{Section:AerodynamicPropertiesCFD}
A finite element (FE) mode has been developed and calibrated in an FE software to obtain mode shapes and frequencies. 
The model is parametrized in a way to easily perform the analysis for sampled section geometries.  
The mode shapes have been identified as vertical bending, lateral bending and torsional modes for the deck considering the reference section. 
However, to automatically identify mode shapes for the sampled cross sections, a Mode Shape Similarity Factor (MSSF) has been used as follows: 
\begin{equation}
	\label{Equation:Mode_Shape_Similarity_Factor}
	\psi_{ij}=\frac{\int_{L}\phi_{i}\left(x\right)\phi_{j}\left(x\right)dx}{\int_{L}\phi_{i}^{2}\left(x\right)dx}\frac{\int_{L}\phi_{i}\left(x\right)\phi_{j}\left(x\right)dx}{\int_{L}\phi_{j}^{2}\left(x\right)dx},
\end{equation}
where $\psi_{ij}$ is known as MSSF. 
Index $i$ is the mode number for the sampled section and index $j$ is for the reference section. 
$\phi$ is the mode shape. 

\begin{figure*}[!htbp]
	\centering
	\includegraphics[scale=0.9,trim=0mm 0mm 4mm 3mm,clip]{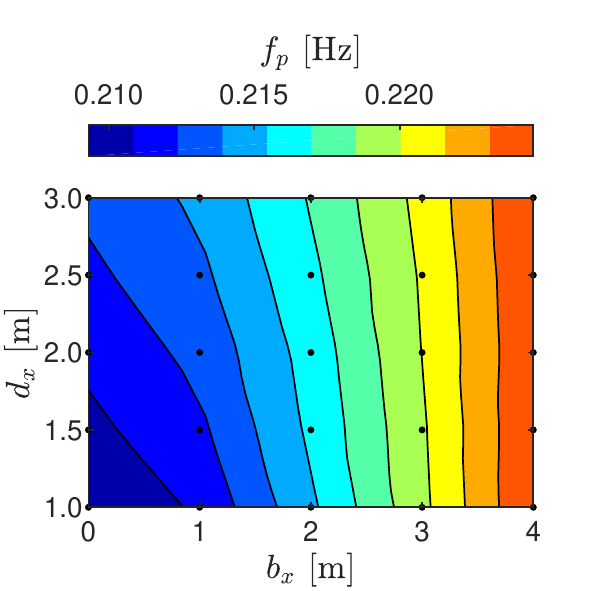}
	\includegraphics[scale=0.9,trim=0mm 0mm 4mm 3mm,clip]{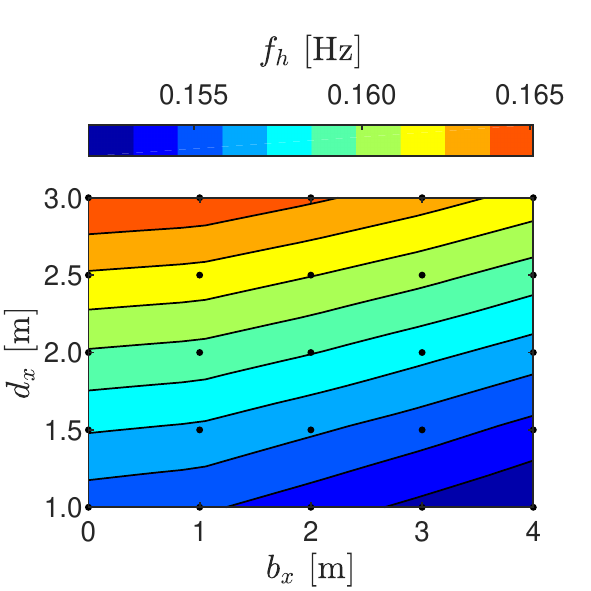}
	\includegraphics[scale=0.9,trim=0mm 0mm 4mm 3mm,clip]{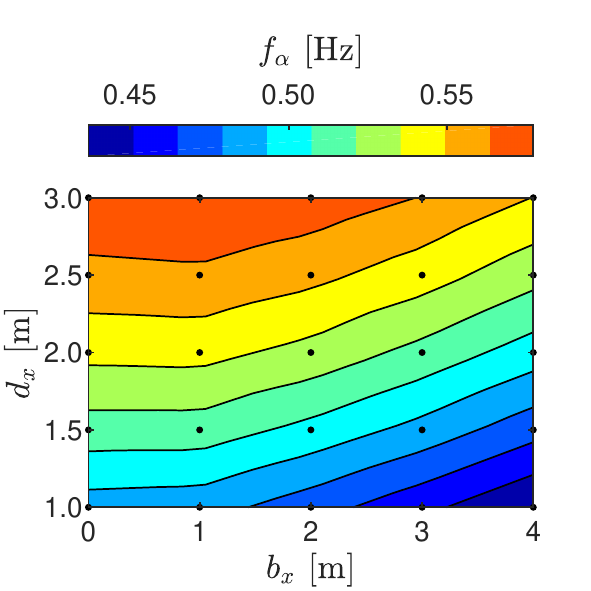}\\
	\includegraphics[scale=0.9,trim=0mm 0mm 4mm 3mm,clip]{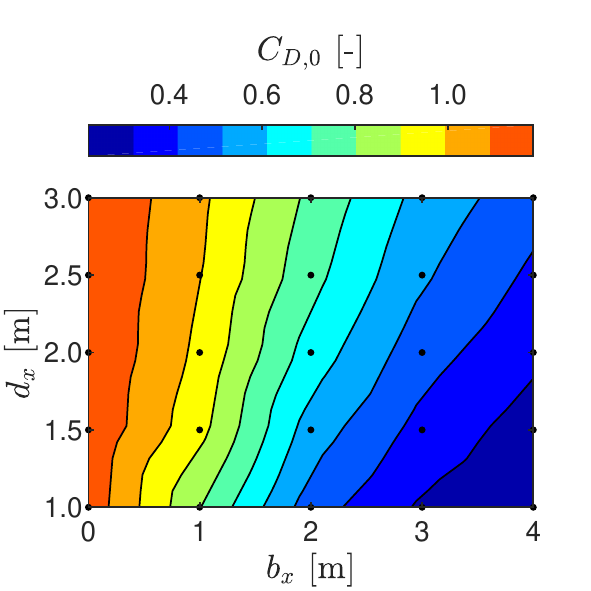}
	\includegraphics[scale=0.9,trim=0mm 0mm 4mm 3mm,clip]{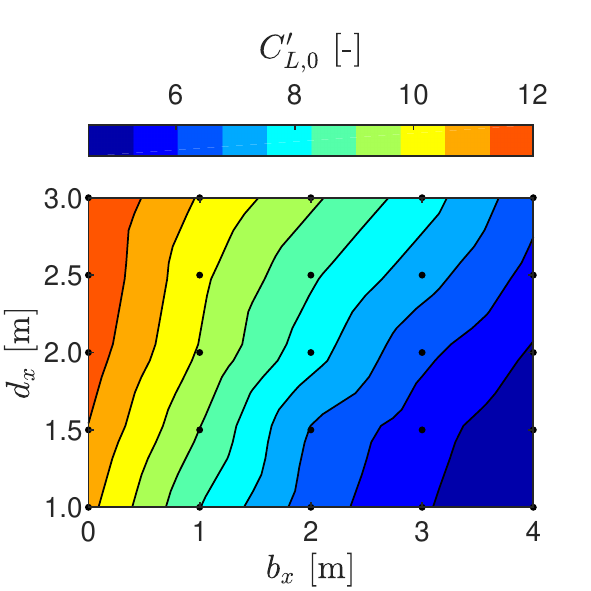}
	\includegraphics[scale=0.9,trim=0mm 0mm 4mm 3mm,clip]{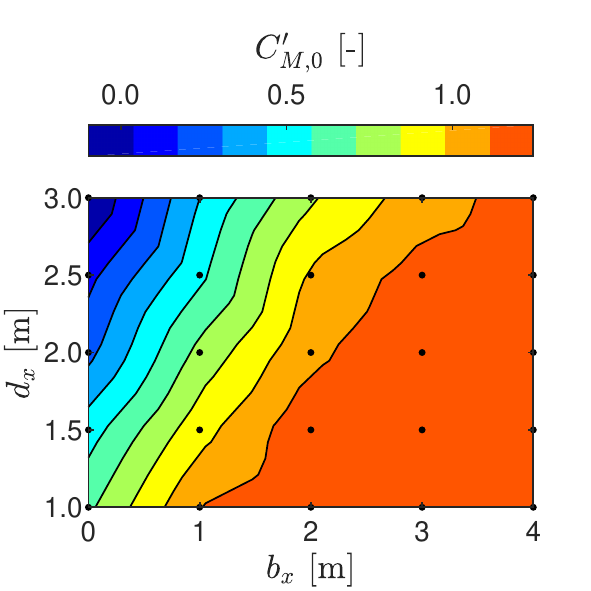}\\
	\includegraphics[scale=0.9,trim=0mm 0mm 4mm 3mm,clip]{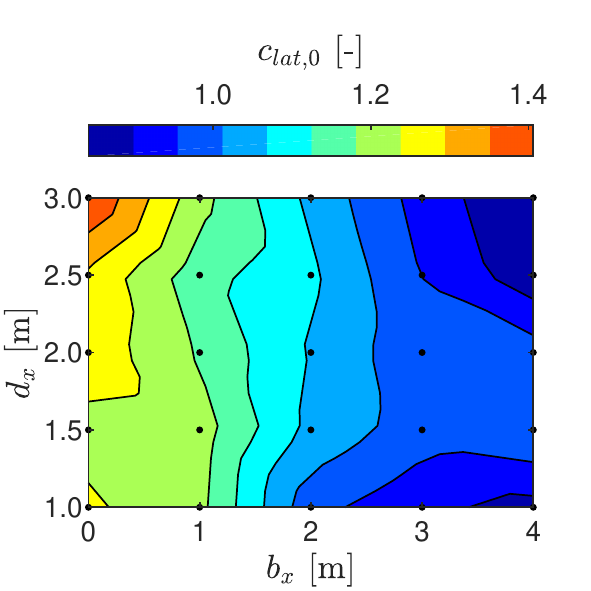}
	\includegraphics[scale=0.9,trim=0mm 0mm 4mm 3mm,clip]{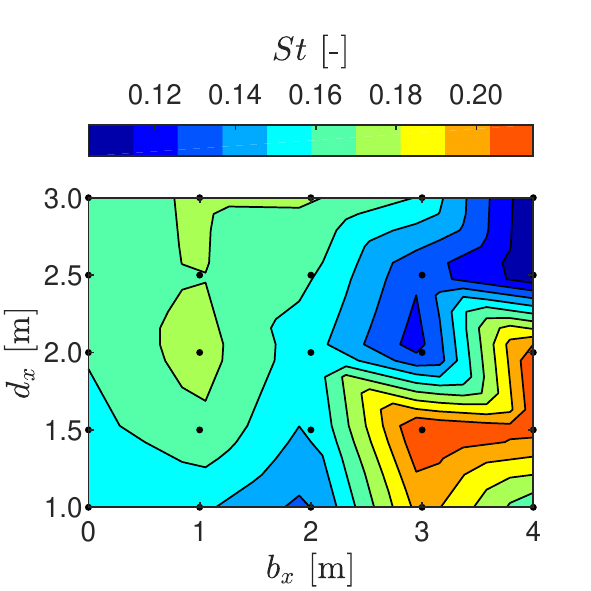}\\
	\includegraphics[scale=0.9,trim=0mm 0mm 4mm 3mm,clip]{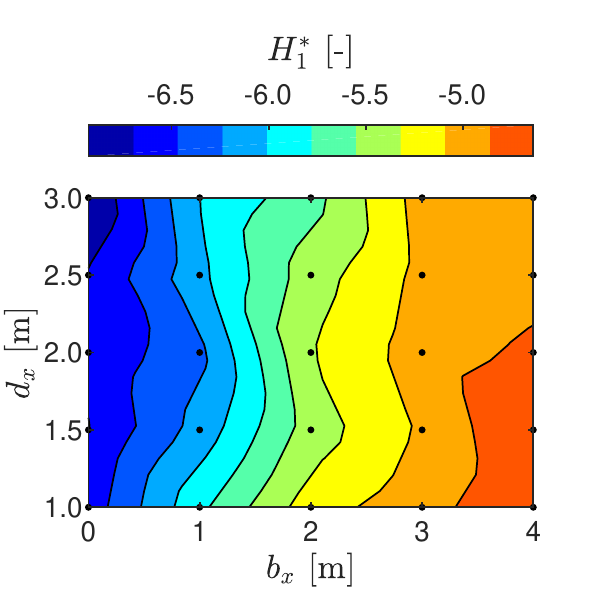}
	\includegraphics[scale=0.9,trim=0mm 0mm 4mm 3mm,clip]{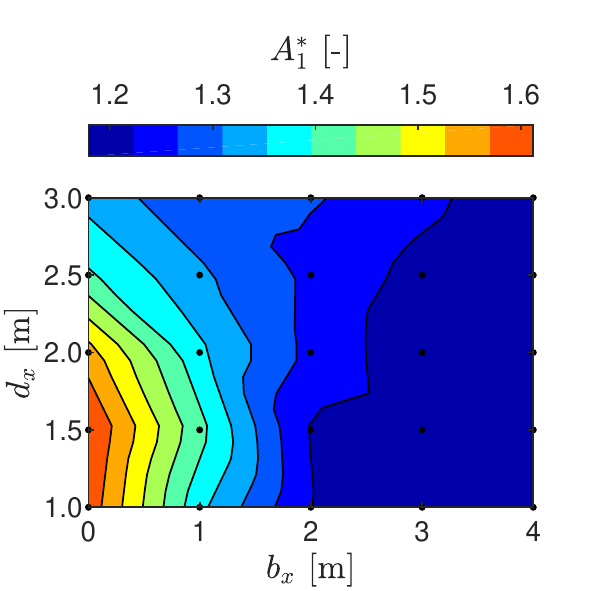}
	\includegraphics[scale=0.9,trim=0mm 0mm 4mm 3mm,clip]{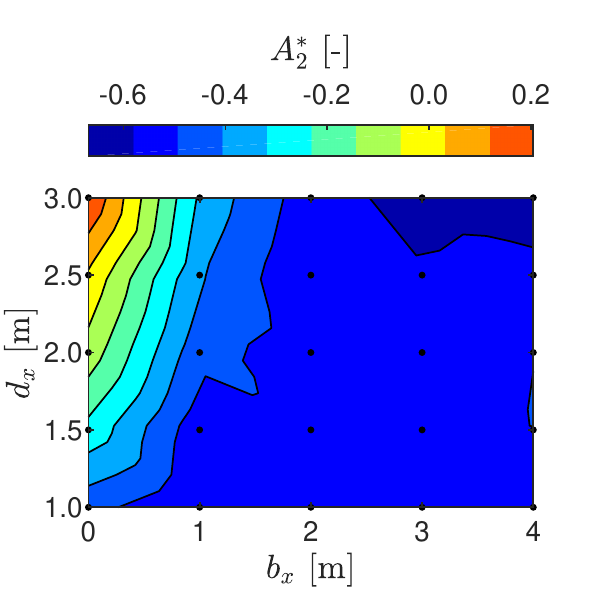}\\
	\caption{Structural and aerodynamic parameters of the deck cross section in the design space. (top to bottom) first natural frequencies; static wind coefficient and slopes at wind angle of attack $\alpha_s=0^\circ$; lateral force coefficient and Strouhal number; aerodynamic derivatives at $v_r=8$.}
	\label{Figure:AerodynamicProperties}
\end{figure*}

This factor compares the computed mode shapes with the reference modes and provides a value between 0 and 1. 
Modes are less similar if the value of the MSSF is close to 0 and are more similar if it is close to 1. 
The identification of mode shapes is required to quantify the sensitivity of frequency due to the change in section geometry parameters. 
Additionally, this information is required for the VIV analysis to identify the vertical bending modes. 
Figure~\ref{Figure:AerodynamicProperties} provides the fundamental frequencies obtained for the samples after the modal analysis. 

\begin{figure}[!htbp]
	\centering
	\includegraphics[scale=0.9,trim=0mm 0mm 0mm 0mm,clip]{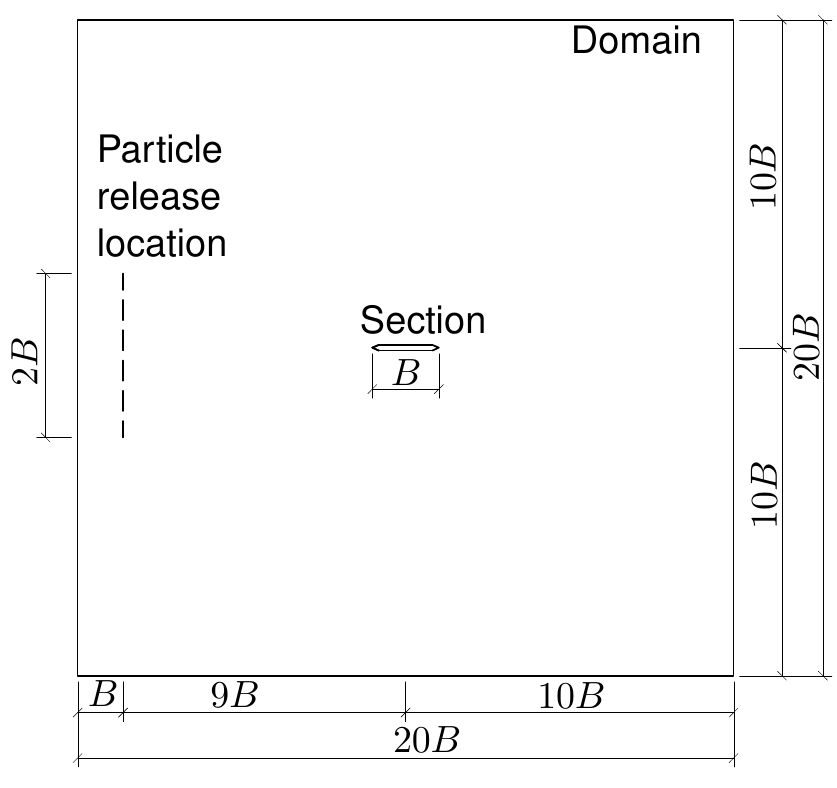}
	\caption{Schematic of CFD simulation domain.}
	\label{Figure:Figure_CFD_Domain}
\end{figure}

\begin{figure*}[!htbp]
	\centering
	\includegraphics[scale=0.70,trim=25mm 10mm 25mm 15mm,clip]{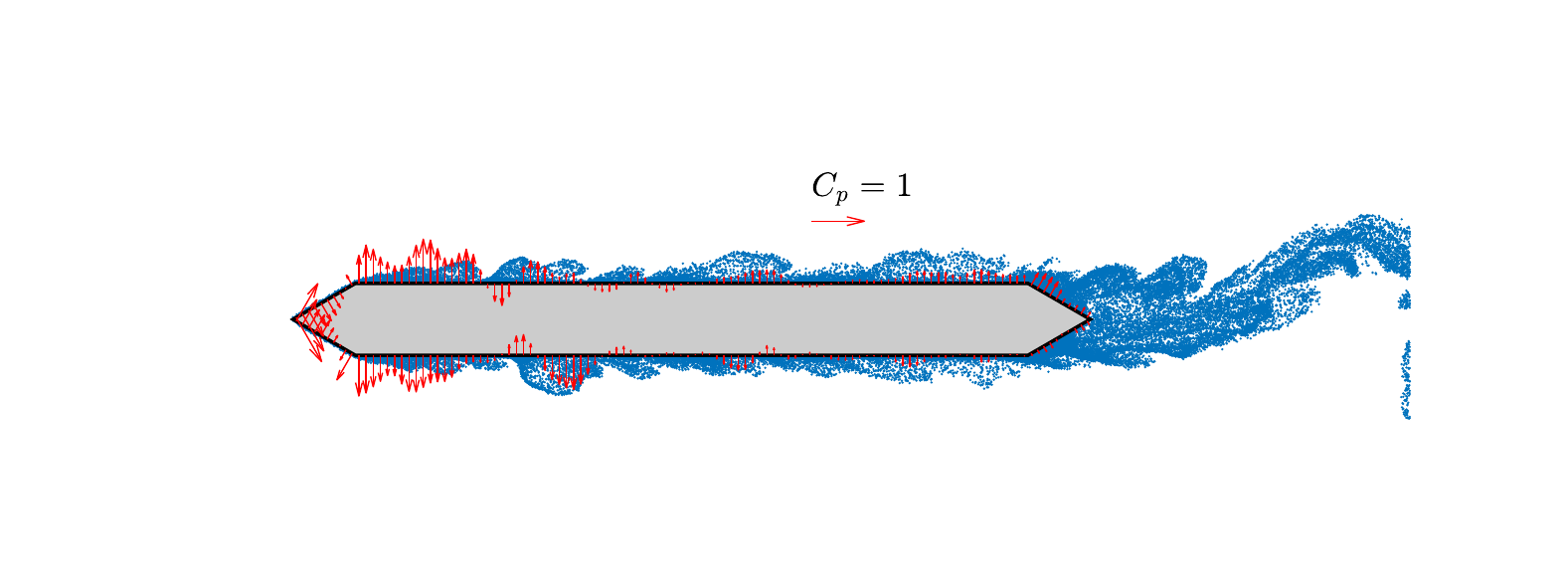}
	\includegraphics[scale=0.70,trim=25mm 10mm 25mm 15mm,clip]{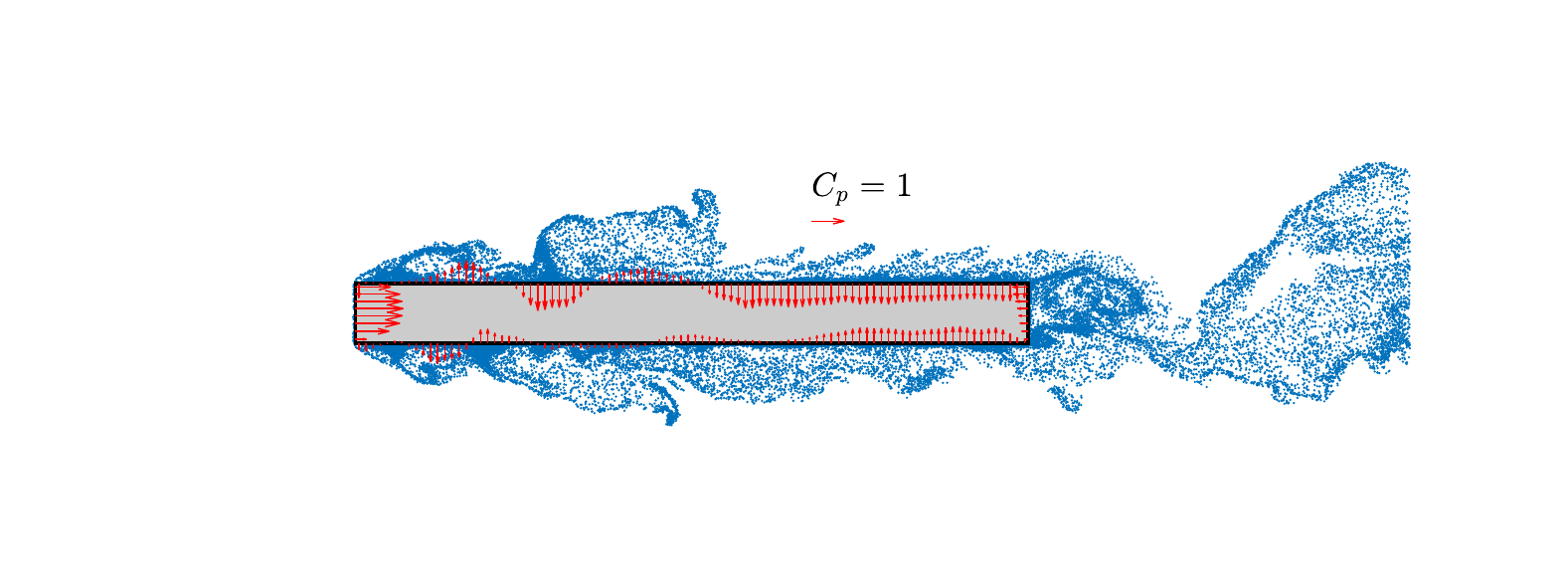}\\
	\includegraphics[scale=0.70,trim=25mm 5mm 25mm 15mm,clip]{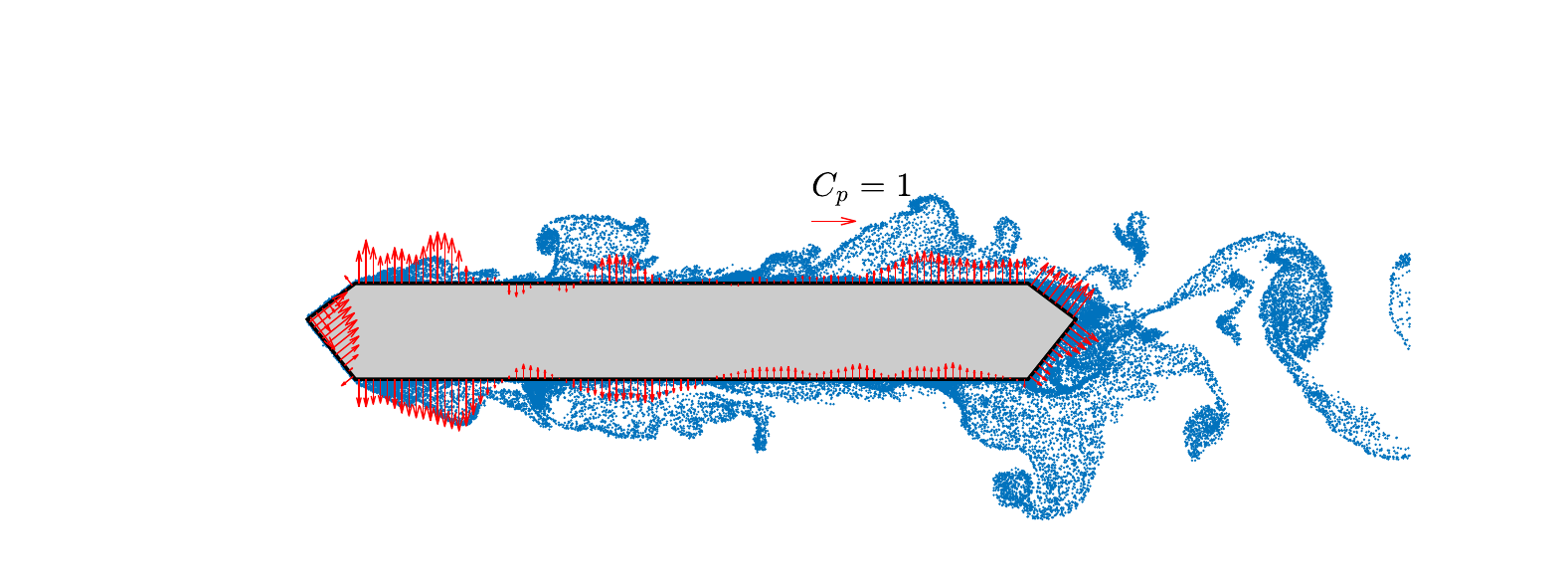}
	\includegraphics[scale=0.70,trim=25mm 5mm 25mm 15mm,clip]{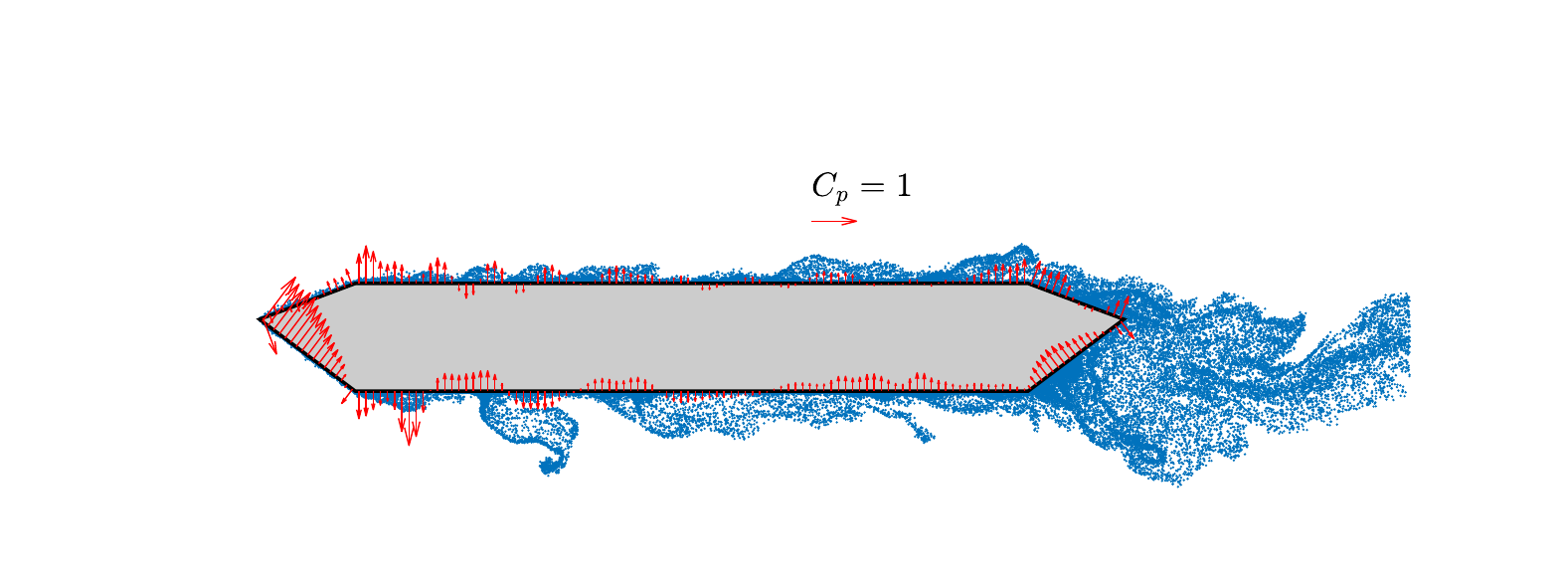}
	\caption{Instantaneous particle map from static simulations for selected samples: (top-left) reference case, (top-right) Sample$\#$5, (bottom-left) Sample$\#$17, (bottom-right) Sample$\#$25, (see also Table~\ref{Table:Mechanical_Properties}). $C_p$ is coefficient of pressure.}
	\label{Figure:ParticleMap}
\end{figure*}

The generated cross section geometries in Section~\ref{Section:DeckShapeParametrization} are modelled in the CFD solver to obtain the aerodynamic properties. 
The numerical simulations are performed to compute static wind coefficients, aerodynamic derivatives and aerodynamic admittance functions. 
The methodology presented in Section~\ref{Section:AerodynamicParameters} has been utilized here for this purpose. 
The domain used in the CFD simulations is presented in Figure~\ref{Figure:Figure_CFD_Domain}. 
The section is placed at the center of the domain. 
Some sampled sections modelled in the CFD have been presented in Figure~\ref{Figure:ParticleMap}. 

The integrated pressure around the section results in the aerodynamic forces in the global degrees of freedman. 
It can be clearly seen in the figure that deeper sections produce larger vortices which in turn produce low frequency vortex shedding forces. 
This results in a potential VIV excitation when the shedding frequency comes close to the natural frequency of the system.

The static wind coefficients are determined by performing simulations under smooth flow at several wind angles of attack i.e. from -10$^\circ$ to 10$^\circ$ with a step of 2$^\circ$. 
The parameters used in the CFD simulations have been presented in Table~\ref{Table:CFD_Parameters}. 
Each simulation is run to account for 200 chords (i.e. $tU/B=200$). 
Whereas, at $\theta=0^\circ$, the simulations were run for 400 chords to have a refined spectrum to identify Strouhal number $St$ (see Eq.~\eqref{Equation:StrouhalNumber}). 
The same simulation was also used for determining the lateral wind coefficient $C_{lat,0}$ (see eq.~\eqref{Equation:LateralForceCoefficient}). 
It was expected that in some cases it would be challenging to obtain a clear peak of the spectral amplitudes of the lift coefficient. 
Therefore the Strouhal number $St$ was determined as a maximum value in a range between 0.05 to 0.16 which is typical for such bridge decks. 

\begin{figure*}[!htbp]
	\centering
	\includegraphics[scale=0.9,trim=15mm 0mm 15mm 0mm,clip]{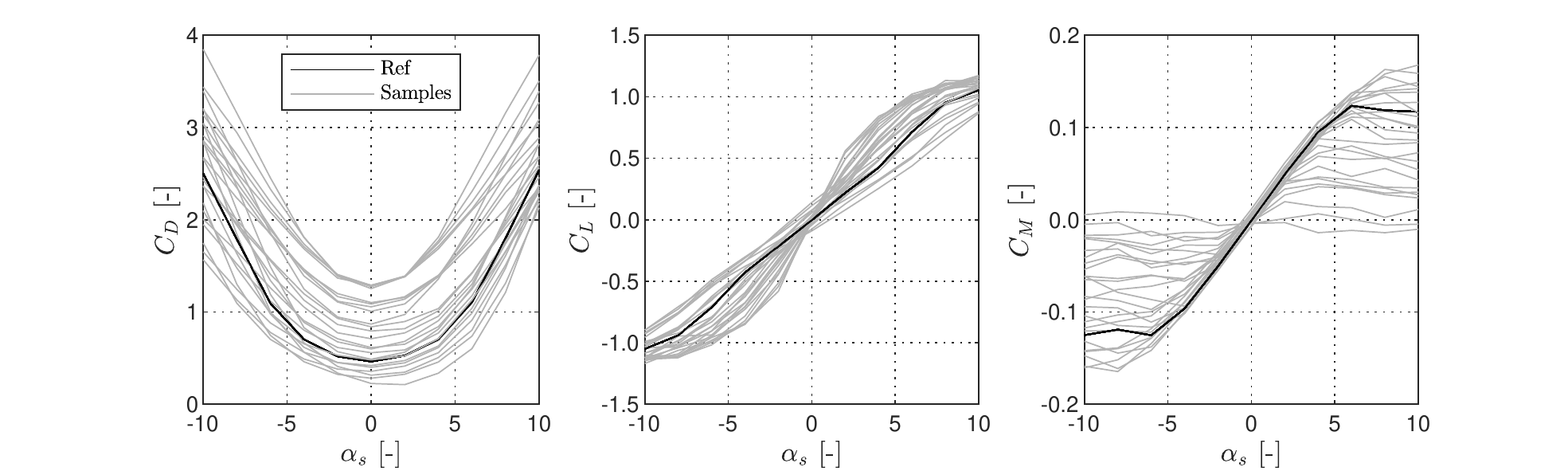}\\
	\includegraphics[scale=0.9,trim=15mm 0mm 15mm 0mm,clip]{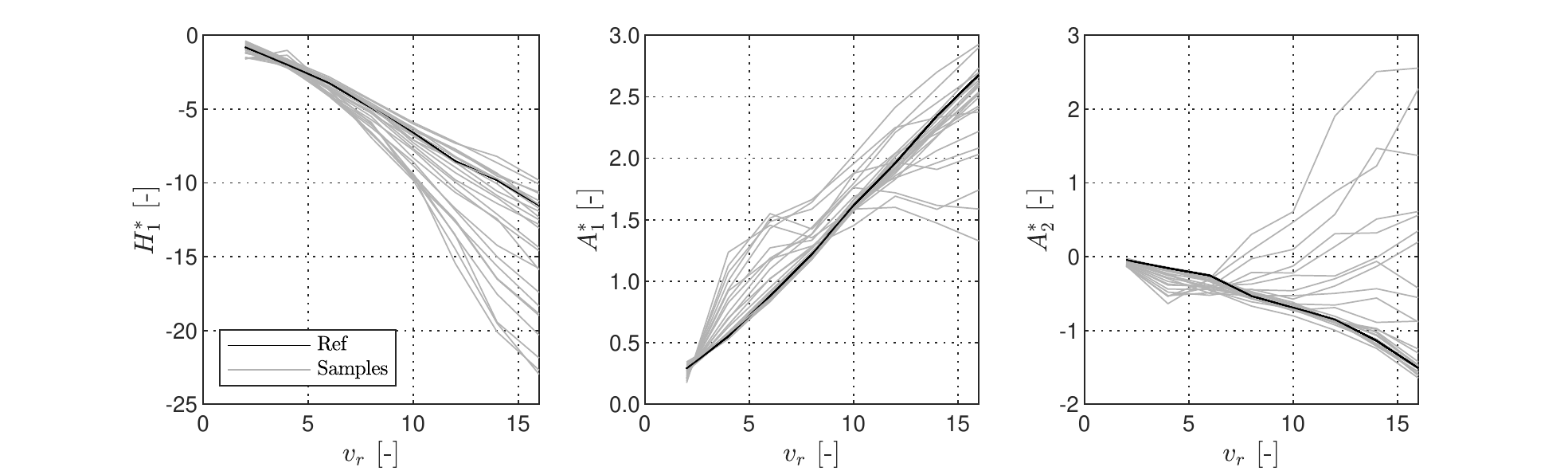}\\
	\includegraphics[scale=0.9,trim=15mm 0mm 15mm 0mm,clip]{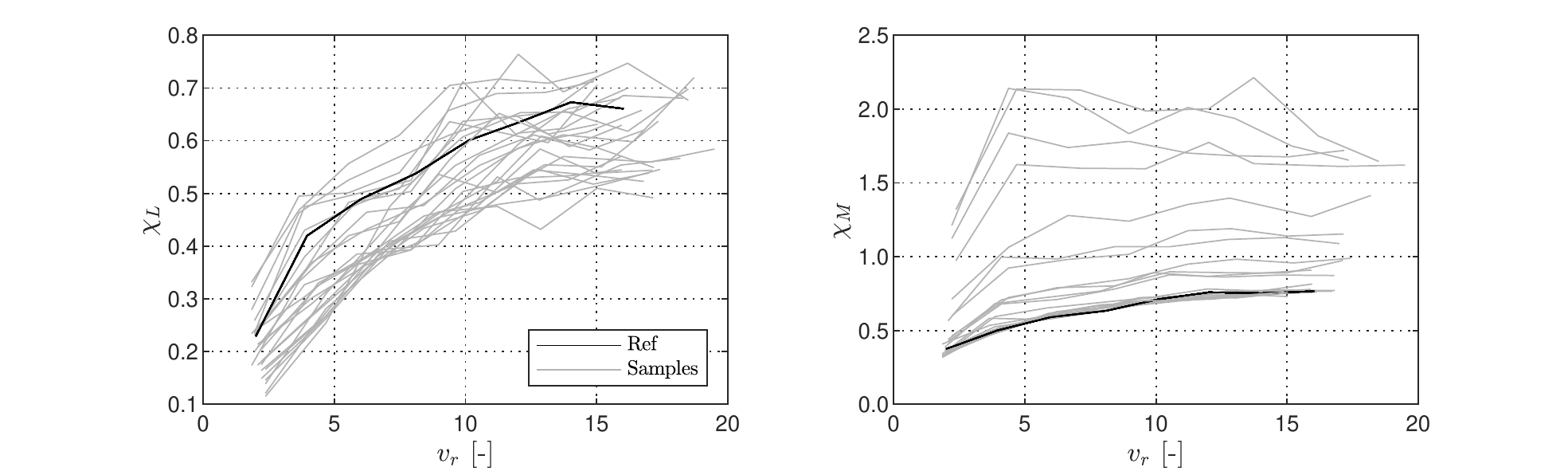}\\
	\caption{Aerodynamic properties of the sampled cross sections: (top) static wind coefficients, (middle) aerodynamic derivatives, (bottom) aerodynamic admittance coefficients.}
	\label{Figure:Step8_StaticWindCoefficients_AerodynamicDerivatives_AerodynamicAdmittance}
\end{figure*}

The flow past the bridge section is simulated to compute time averaged wind coefficients of drag $C_D$, lift $C_L$ and moment $C_M$ for various angle of attacks $\alpha_s$. 
The static wind coefficients are shown in Figure\ref{Figure:Step8_StaticWindCoefficients_AerodynamicDerivatives_AerodynamicAdmittance} with respect to the wind angle of attack $\alpha_s$. 
Since the reference cross-section is symmetric about the horizontal axis, the absolute values of the lift and the moment coefficients are close to zero which may not be the case for other sections. 
The slope of the lift and the moment coefficients are obtained for the values $\alpha_s=\pm2^\circ$. 
It can be seen from the figure that the linearity in the lift and moment slopes is maintained between $\pm4^\circ$ for most cases. 
Figure~\ref{Figure:AerodynamicProperties} summarizes the static wind coefficients and their slopes at $\alpha_s=0^\circ$.

\begin{table}[!htbp]
	\caption{Parameters of the CFD simulations.}
	\label{Table:CFD_Parameters}
	\centering
	\begin{tabularx}{0.5\textwidth}{L{0.62}R{0.38}}
		\hline
		\multicolumn{2}{l}{Common parameters}\\
		\hline
		Wind speed $U_\infty$ [m/s] 		& 10\\
		Reduced time step $\Delta t^*=\Delta tU/B$	& 0.1\\
		No. of panels  						& 200 - 250\\ 
		Panel length $\Delta s/B$			& 0.009\\ 
		Reynolds number $Re$ 				& 1.0$\times$ 10$^5$\\
		Time step $\Delta t$ [s]			& 0.3\\
		No. of particles			        & 130,000-160,000\\
		\hline
	\end{tabularx}
	\begin{tabularx}{0.5\textwidth}{L{0.7}R{0.3}}
		\multicolumn{2}{l}{Static (11 simulation per sample)}\\
		\hline
		Wind attack angle $\theta$ [$^\circ$]       & -10:2:10\\
		Total time for coefficients:  $tU/B$ & 200 \\ 
		Total time for $St$, $c_{lat,0}$: $tU/B$    & 400\\
		\hline
	\end{tabularx}
	\begin{tabularx}{0.5\textwidth}{L{0.7}R{0.3}}
		\multicolumn{2}{l}{Forced vibration (16 simulation per sample)}\\
		\hline
		Reduced speed $v_r$ 		        & 2:2:16\\
		No. of cycles $t/T_o$				& 15\\
		Rotational forced amplitude $\alpha_o$ [$^\circ$]	& 3\\
		Vertical forced amplitude $h_o$ [m] & $UT_o \tan^{-1}(\alpha_o)$ \\
		\hline
	\end{tabularx}
	\begin{tabularx}{0.5\textwidth}{L{0.7}R{0.3}}
		\multicolumn{2}{l}{Admittance (8 simulation per sample)}\\
		\hline
		Vertical intensity $I_w$ [$\%$] 	& 6\\
		Reduced velocities $v_r$ 	        & 2:2:16 \\
		\hline
	\end{tabularx}
\end{table}

The aerodynamic derivatives are determined using the forced vibration simulations in a uniform flow. 
Motion-induced aerodynamic forces for the cross sections are determined from CFD simulations on a 2D section model that is forced to oscillate with a pre-defined harmonic displacement. 
For this purpose, the forcing properties used are given in Table~\ref{Table:CFD_Parameters}. 
The resulting lift force $F_L$ and moment $F_M$ time histories are used to compute the aerodynamic derivatives in a least-squares sense as mentioned in Section~\ref{Section:AerodynamicParameters}. 
For each reduced velocity $v_r$, the simulation time corresponding to 15 periods was established (i.e. $t/T_o=15$). 
This ensures sufficient accuracy for the least-squares fit. 
The aerodynamic derivatives for the section are shown in Figure~\ref{Figure:Step8_StaticWindCoefficients_AerodynamicDerivatives_AerodynamicAdmittance} and Figure~\ref{Figure:AerodynamicProperties}. 

Aerodynamic admittance functions are computed form static simulations with oncoming deterministic gust. 
For this purpose, the particles are released on the upstream side of the section containing the vorticity tuned to a specific reduced speed and amplitude. 
The resulting lift and moment time traces as well as the wind speed components at the section location are used to compute aerodynamic admittance functions. 
The simulation parameters have been provided in Table~\ref{Table:CFD_Parameters}. 
The computed aerodynamic admittance functions are shown in Figure~\ref{Figure:Step8_StaticWindCoefficients_AerodynamicDerivatives_AerodynamicAdmittance}.

\subsection{Aeroelastic analysis}
\label{Section:Aeroelastic analysis}
Aeroelastic analysis is performed for each sample as explained in Section~\ref{Section:SemiAnalyticalModels} to obtain buffeting response, flutter limits and VIV amplitudes. 

\begin{figure*}[!htbp]
	\centering
	\includegraphics[scale=0.9,trim=15mm 0mm 15mm 0mm,clip]{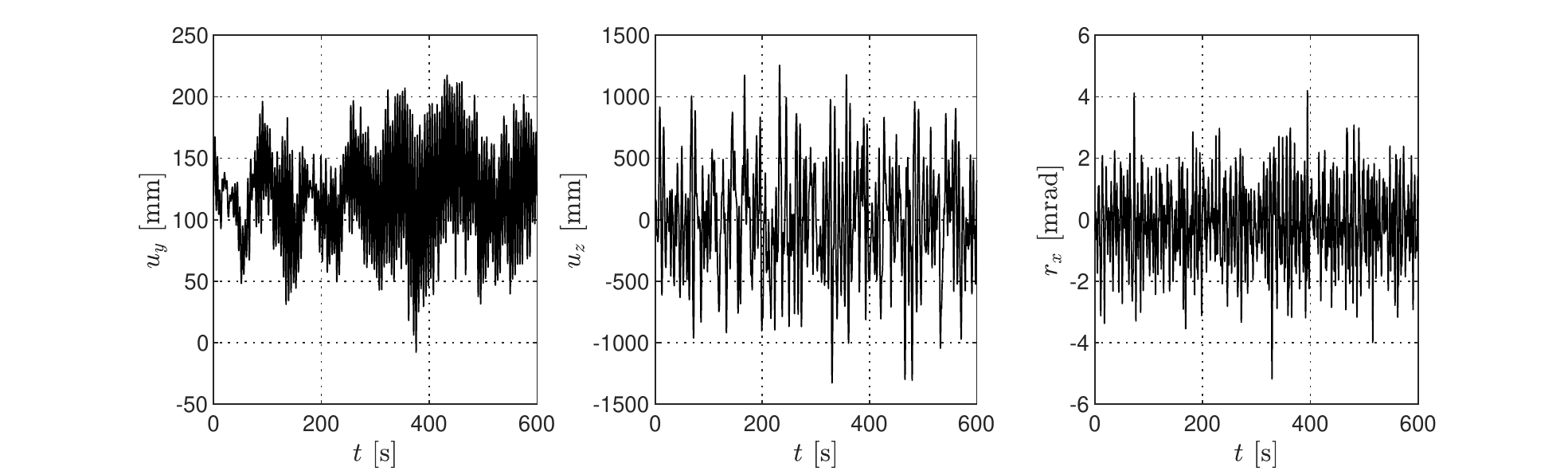}\\
	\includegraphics[scale=0.9,trim=15mm 0mm 15mm 0mm,clip]{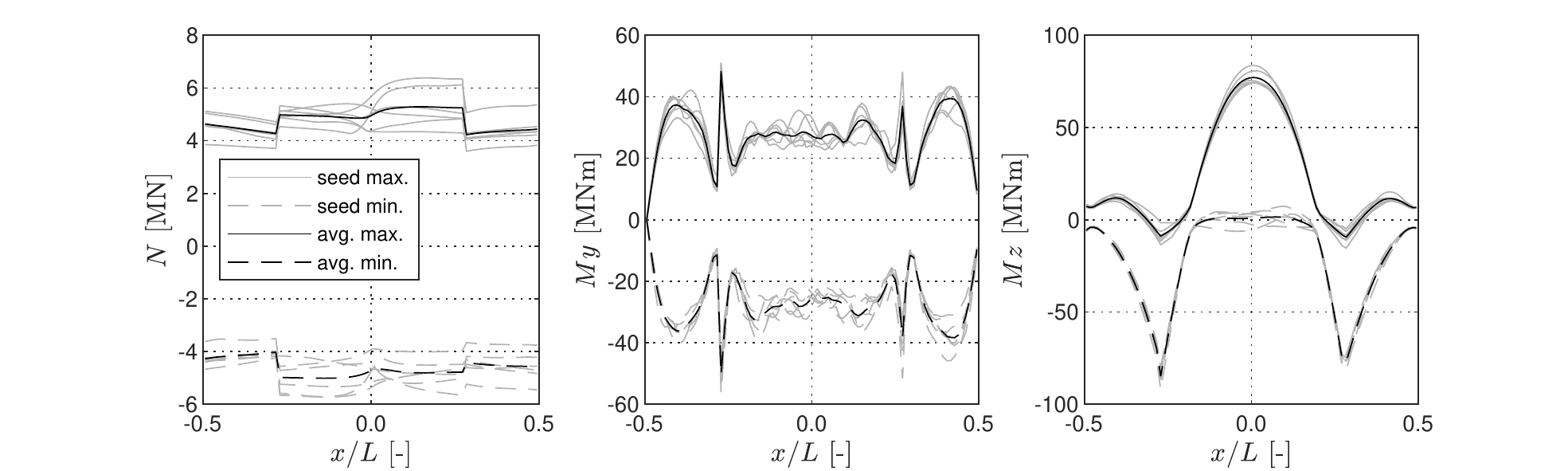}
	\caption{Buffeting response analysis for the reference case: (top) response time histories at the central node $x/L=0$ (bottom) envelope of forces.}
	\label{Figure:BuffetingResponseDeckInternalForces}
\end{figure*}

\begin{figure*}[!htbp]
	\centering
	\includegraphics[scale=0.9,trim=0mm 0mm 4mm 3mm,clip]{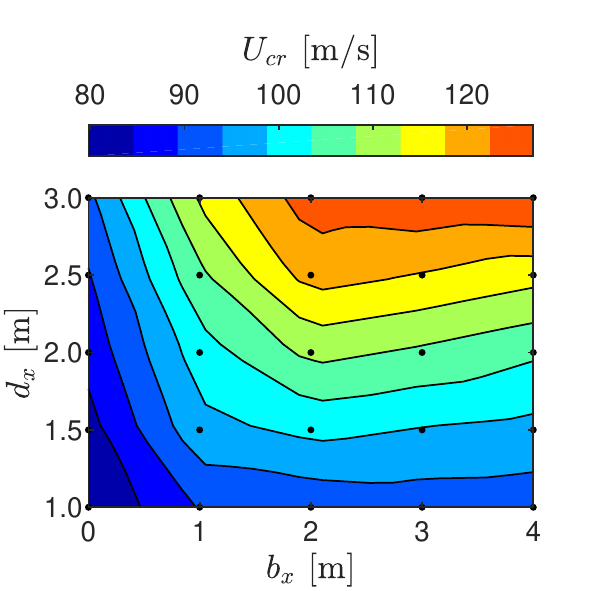}
	\includegraphics[scale=0.9,trim=0mm 0mm 4mm 3mm,clip]{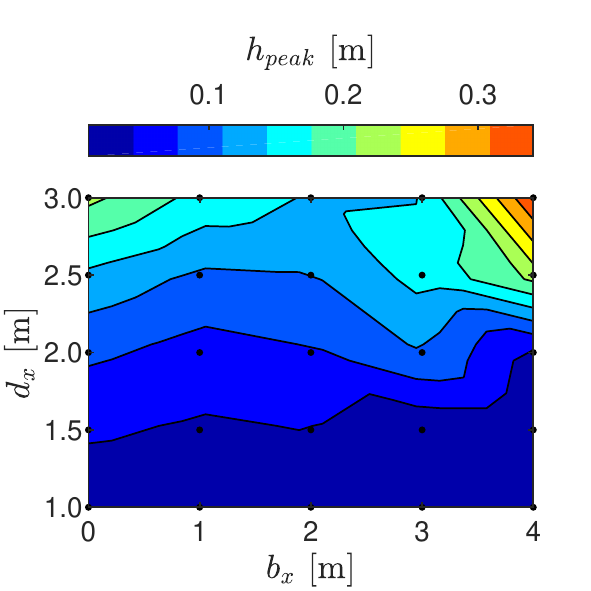}
	\includegraphics[scale=0.9,trim=0mm 0mm 4mm 3mm,clip]{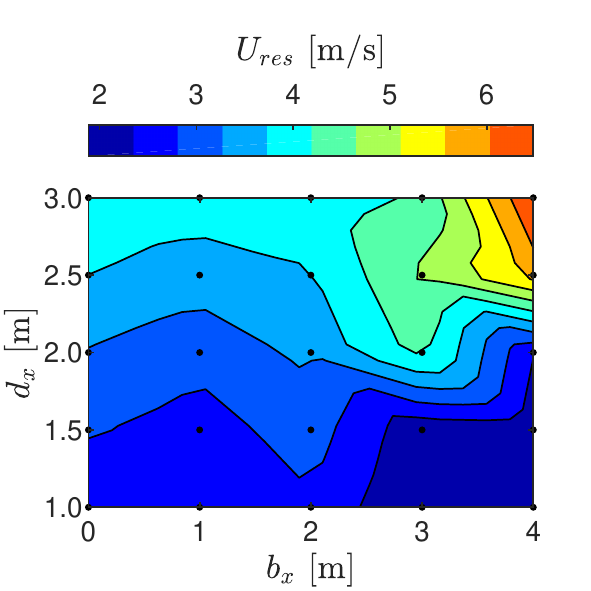}\\
	\includegraphics[scale=0.9,trim=0mm 0mm 4mm 3mm,clip]{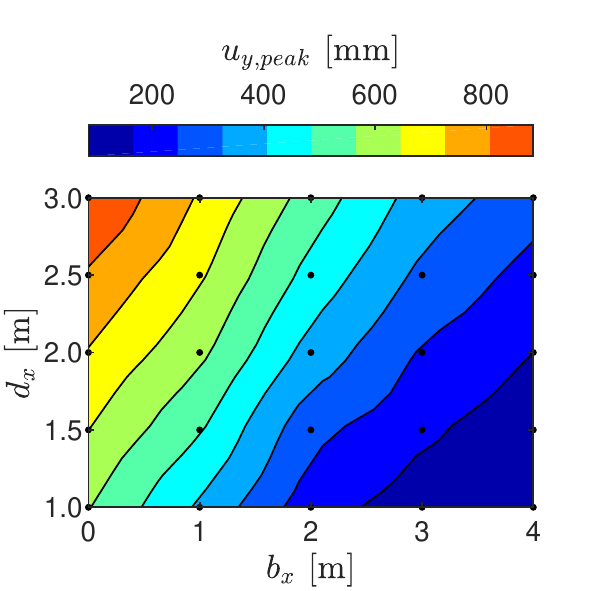}
	\includegraphics[scale=0.9,trim=0mm 0mm 4mm 3mm,clip]{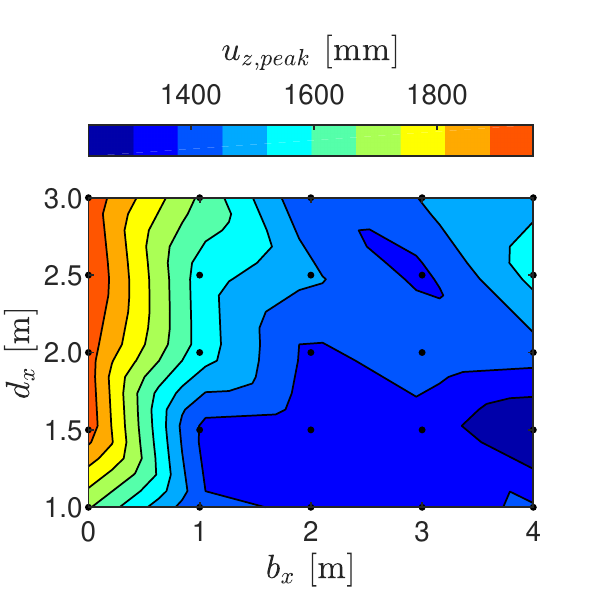}
	\includegraphics[scale=0.9,trim=0mm 0mm 4mm 3mm,clip]{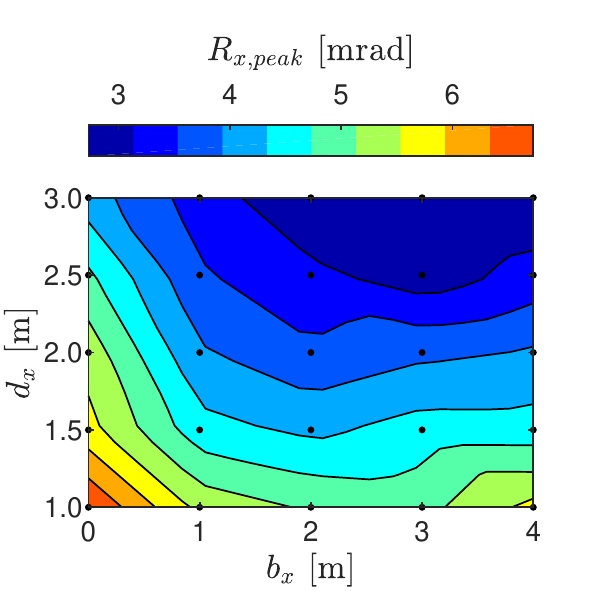}\\
	\includegraphics[scale=0.9,trim=0mm 0mm 4mm 3mm,clip]{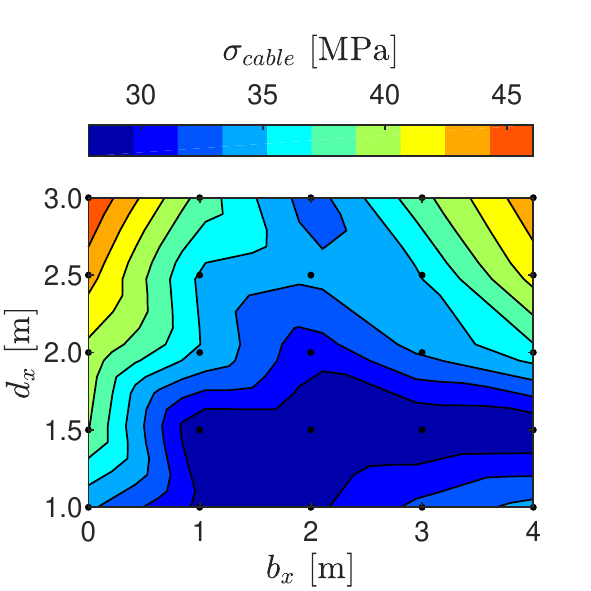}
	\includegraphics[scale=0.9,trim=0mm 0mm 4mm 3mm,clip]{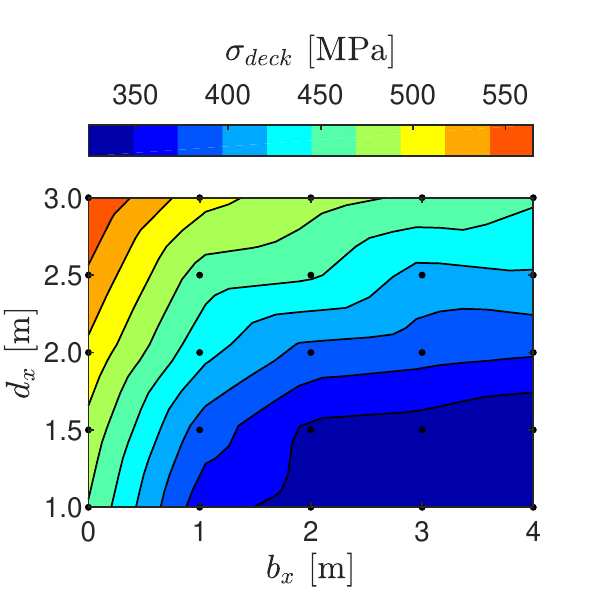}\\
	\caption{Aeroelastic analysis: (top) flutter limit $U_{cr}$, peak amplitudes due to VIV $h_{peak}$, resonant wind speed for VIV $U_{res}$ of the first mode, (middle) peak deck amplitudes from buffeting analysis $u_{y,peak}$, $u_{z,peak}$, $R_{x,peak}$, (bottom) cable and deck peak stresses from buffeting analysis.}
	\label{Figure:FlutterVIVBuffeting}
\end{figure*}

Buffeting analysis is performed in the time domain by generating wind time histories and carrying out the dynamic analysis in the mode space. 
For this purpose, wind characteristics presented in Table~\ref{Table:Wind_characteristics} have been used. 
These wind characteristics have been obtained from \cite{Publication_IIIR} and Eurocode \cite{Eurocode2004}. 
First 50 modes have been used in the analysis. 
The structure is exposed to the six 10-min wind time histories and the results are obtained which are then averaged. 
Some sampled response time histories and envelopes of forces have been shown in Figure~\ref{Figure:BuffetingResponseDeckInternalForces}. 
The buffeting forces are combined with the dead and live loads to compute the overall stresses in the deck section. 
The peak values of stresses along the whole length of the deck are monitored and recorded. The stresses in the main cable are also checked. 
Figure~\ref{Figure:FlutterVIVBuffeting} provided the peak deck response and stress in the deck and the main cable. 
As compared to flutter and VIV, the buffeting analysis is the most computationally intensive. 
It involves setting up the input files, performing the time history analysis and then post processing the results by generating envelopes of nodal response and member forces six times for each sample. 
It was made sure that the data transfer between different components of the framework is efficient without the loss of useful information and avoiding unnecessary steps or repetitive steps. 

\begin{table}[!htbp]
	\caption{Wind characteristics used in the buffeting analysis (based on Eurocode \cite{Eurocode2004}).}
	\label{Table:Wind_characteristics}
	\centering
	\begin{tabularx}{0.5\textwidth}{L{0.7}R{0.3}}
		\hline
		Mean wind speed [m/s]					& 52.0\\
		Longitudinal turbulent intensity [-] 	& 0.104\\
		Vertical turbulent intensity [-] 		& 0.058\\
		Longitudinal length scale [m]			& 169\\
		Vertical length scale [m]				& 56\\
		Spectrum of wind fluctuations 			& von Karman\\
		\hline
	\end{tabularx}
\end{table}

\begin{figure*}[!htbp]
	\centering
	\includegraphics[scale=0.9,trim=15mm 0mm 15mm 0mm,clip]{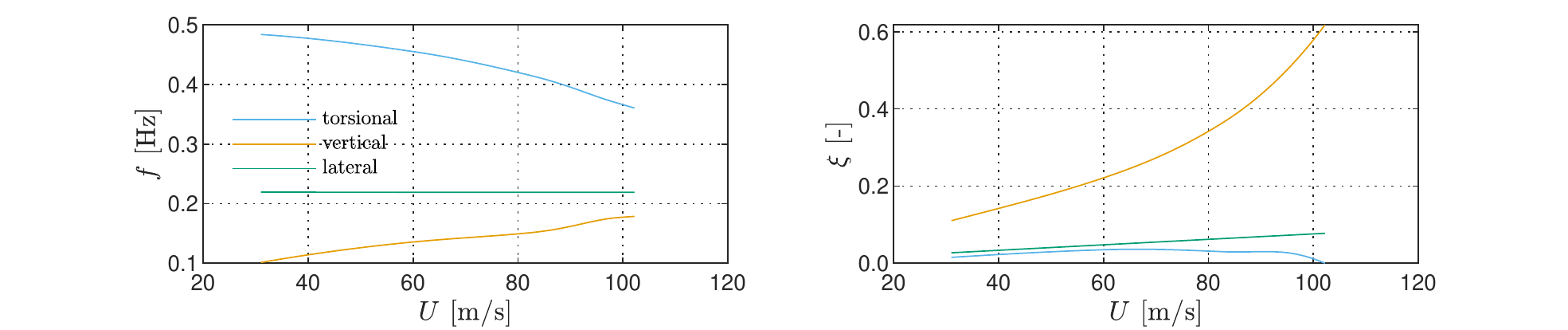}
	\caption{Flutter analysis for the reference case: prediction of flutter limit from eigenvalue analysis (left)~frequencies, (right)~damping ratios.}
	\label{Figure:FrequencyDamping}
\end{figure*}

The flutter analysis is performed to predict flutter limit using fundamental vertical bending, lateral bending and torsional modes by performing eigenvalue analysis using multimode flutter solution \cite{Abbas2016P}. 
This analysis utilizes aerodynamic derivatives for each section geometry and computes the frequency and damping ratio for the modes considered at different wind speeds. 
The frequency and damping ratio along wind speed have been presented in Figure~\ref{Figure:FrequencyDamping} for the reference section. 
It can be seen that the damping ratio becomes zero at around 101~m/s indicating the flutter onset. 
The figure also shows that the torsional frequency decreases and the vertical bending frequency increases indicate mode coupling. 
The lateral frequency is however, not affected indicating that this mode is not contributing in the flutter response. 
The torsional flutter limit is also checked based on the eq.~\eqref{Equation:TorsaionalFLutter}. 
For some rectangular and deeper section geometries, the torsional flutter is observed. 
This can be seen in Figure~\ref{Figure:AerodynamicProperties} where the aerodynamic derivative $A_2^*$ becomes positive at a certain reduced speed $v_r$. 
The analysis only required a few minutes for all the sample geometries due to the advantage of using frequency domain. 

\begin{figure*}[!htbp]
	\centering
	\includegraphics[scale=0.9,trim=0mm 0mm 0mm 0mm,clip]{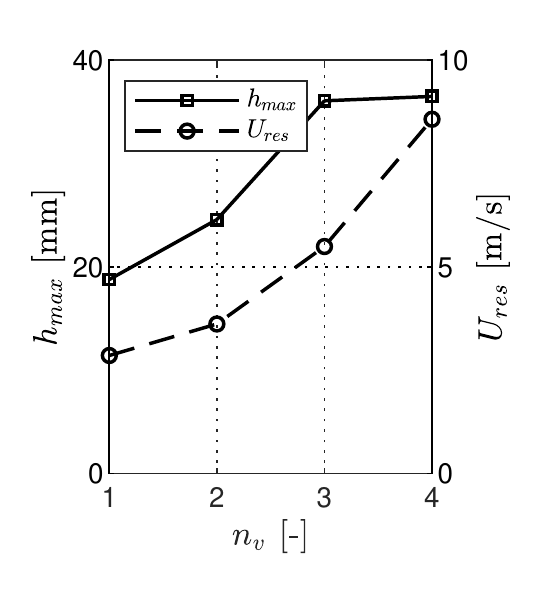}
	\includegraphics[scale=0.9,trim=0mm 0mm 0mm 0mm,clip]{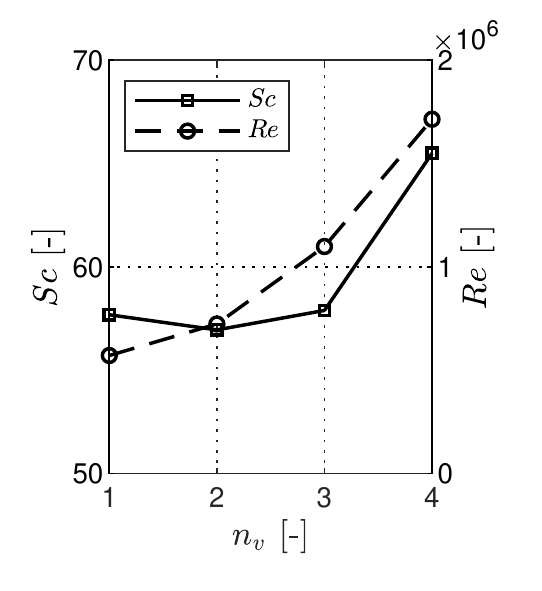}
	\includegraphics[scale=0.9,trim=0mm 0mm 0mm 0mm,clip]{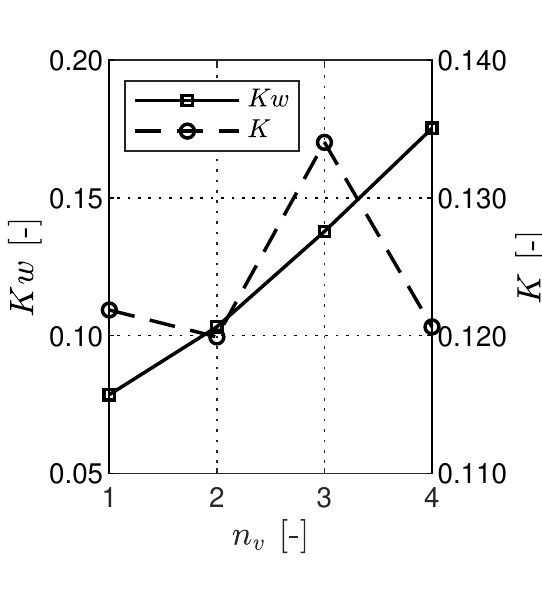}
	\caption{Vortex-induced vibration analysis for the reference case: parameters for different modes.}
	\label{Figure:VIVparameters}
\end{figure*}

The VIV amplitudes have been computed based on the procedure presented in Eurocode \cite{Eurocode2004}. 
This method has been developed by Ruscheweyh \cite{Sockel1994B} and is widely regarded as realistic and rational for VIV excitation of slender structures. 
The method is summarized in Section~\ref{Section:VortexInducedVibrations}. 
The VIV analysis is performed considering wind acting on the deck only.
This has been done for the first 4 vertical bending modes of the deck. 
This covers all modes that can be excited by resonance with vortex shedding within the design wind speed range. 
The Strouhal number $St$ for the bridge section is obtained from the static CFD simulations at $\alpha_s=0^\circ$. 
This was achieved by the dominant frequency of the lift coefficient time history. 
The lateral force coefficient $c_{lat,0}$ has been also obtained from static simulations by obtaining the standard deviation of lift coefficient normalized with respect to the depth $D$. 
The Scruton number $Sc$ is then calculated from modal properties. 
Maximum response amplitudes $y_{F,max}$ are computed based on eq.~\eqref{Equation:EurocodeVIV}. 
Aerodynamic damping is considered in the VIV analysis. 
The peak amplitude and and other related parameters are presented in Figure~\ref{Figure:VIVparameters}.

\subsection{Response surface generation}
\label{Section:Response surface generation}
The commuted quantities of interest for all the section geometries have been utilized to generate response surfaces using MLS as explained in Section~\ref{Section:ResponseSurface}. 
These surrogate models require only a fraction of time to generate the intended output. 
These are then used in the optimization process explained in Section~\ref{Section:AerodynamicOptimization}. 
The input used to generate the response surface are the section geometry parameters $d_x$ and $b_x$. 
The response surfaces are build for the flutter analysis to predict flutter onset for the geometries. 
For the VIV analysis, peak amplitudes have been considered. 
Peak response amplitudes of the deck and the stresses in deck and main cable have been taken to account for buffeting analysis. 
All these required quantities are presented in Figure~\ref{Figure:FlutterVIVBuffeting}. 

\subsection{Sensitivity due to input parameters}
The influence of input parameters on the target quantities has been investigated. 
The structural parameters are most important for the dynamic analysis and directly influence the aeroelastic response. 
The lateral frequency $f_p$ is not much affected by changing the depth $d_x$; however, it increases 6\% by increasing the width $b_x$ from minimum to maximum in the design space (see Figure~\ref{Figure:AerodynamicProperties}). 
The effect on the bending frequency $f_h$ and the torsional frequency $f_\alpha$ is opposite to the previous case. 
The change is generally less affected by the depth $d_x$ and is more by the width $b_x$. 
No significant change has been observed in the mode shapes (see Figure~\ref{Figure:Step5_ModeShapes}) due to changing the deck shape as the stiffness of the whole deck is constant along the length. 

The aerodynamic parameters are highly dependent on the geometric shape. 
Figure~\ref{Figure:AerodynamicProperties} shows the static wind coefficient of drag $C_{D,0}$ and the slopes of lift $C'_{L,0}$ and moment $C'_{M,0}$ coefficients. 
The coefficient of drag $C_{D}$ directly results in the overall increase of drag force on the structure due to increase in the depth of the section. 
However, a streamlined section with a smaller depth to width ratio offers lesser drag. 
For the same depth (at $d_x=1.0$), when the fairings are introduced, a significant reduction in drag is observed of up to 78\% (at $b_x=4$). 
For the selected design space, the drag reduces to 0.34 from 1.53 which is quite a significant reduction; 
However, the drag alone is not the most derisive factor to choose a cross sectional shape. 
The section must be able to perform against other aeroelastic phenomena. 
Relatively small values of lift $C_{L,0}$ and moment $C_{M,0}$ coefficients are found at $\alpha_s=0^\circ$ for  the cross sections symmetrical about the horizontal axis. 
However, the slope of the lift $C'_{L,0}$ and moment coefficients $C'_{M,0}$ more important and contribute significantly to the aeroelastic behaviour. 

The lateral wind coefficient contributes to the VIV behaviour. 
It is large for sections with a smaller aspect ratio since a deeper section sheds larger vertices in turns generating larger lift force. 
The effect of geometric shape is somewhat similar to that of drag coefficient $C_D$. 
It increases with increasing the depth but for a given depth it reduces when fairings are introduced.
The Strouhal number does not show a clear trend for the most part of the design domain; however, it tends to increase for a shallower and wider section. 

The aerodynamic derivatives $H_1^*$, $A_1^*$ and $A_2^*$ are considered significantly important as compared to the other derivatives. 
$H_1^*$ is the lift force due to vertical oscillations of the section. 
This is related to the single degree of freedom galloping instability which occurs when the value of $H_1^*$ becomes positive. 
Figure~\ref{Figure:Step8_StaticWindCoefficients_AerodynamicDerivatives_AerodynamicAdmittance} shows that the aerodynamic derivative $H_1^*$ for all the cross section geometries is negative for all $v_r$ range indicating a stable behaviour against galloping. 
The aerodynamic derivative $A_2^*$ is the moment due to torsional motion and is associated to torsional instability. 
Figure~\ref{Figure:Step8_StaticWindCoefficients_AerodynamicDerivatives_AerodynamicAdmittance} indicates that some samples have a positive $A_2^*$ (from reduced speed $v_r>7$) leading to torsional flutter. 
Figure~\ref{Figure:AerodynamicProperties} depicts aerodynamic derivatives at a reduced speed $v_r=8$ which is found to be close to the aeroelastic instability limit for the reference case. 
Fore some cases, $A_2^*$ is positive indicating the onset of torsional flutter limit. 
This happened for deeper cross sections which are less streamlined. 
Commonly, H-shape and deep rectangular sections are usually prone to torsional flutter \cite{AbbasReview2017J}. 
The aerodynamic derivatives presented in the figure are the also essential for the coupled flutter (see also \cite{AbbasProbabilistic2015J}). 
Coupled flutter can happen even if $A_2^*$ and $H_1^*$ are negative \cite{AbbasReview2017J}. 
Coupled flutter is more critical for a section with smaller width. 
It is also observed that the coupled flutter becomes critical for shallower sections as these sections offer smaller torsional resistance and have low torsional frequency. 
The flutter directly limit depends on the ratio of torsional frequency to bending frequency. 
Figure~\ref{Figure:FlutterVIVBuffeting} indicates that a shallow (with smaller $d_x$) and narrow (with smaller $b_x$) section is more critical for the coupled flutter. 
Whereas a wider section performs much better. 
There is an increase of flutter limit of up to 65\% in the selected design domain between different section geometries. 

The VIV analysis indicates that the oscillation amplitudes are larger for higher vertical bending modes (see Figure~\ref{Figure:VIVparameters}). 
This is mainly due to the larger contribution of mode shape factor $K$ and the Scruton number $Sc$. 
The modes with very high frequency are not prone to the resonance as the resonance wind speed is much higher than the design wind speed of the structure, therefore these modes are not expected to create VIV oscillations. 
The VIV is much more critical for deeper sections. 
The shallower sections shed smaller vertices and thus producing smaller lift forces on the section which creates resonance oscillations. 
The VIV response is not much affected by increasing the section with $b_x$ for a given depth $d_x$. 
Figure~\ref{Figure:VIVparameters} indicates that generally the resonant wind speed is higher for the deeper sections and smaller for shallower sections. 
However, the resonant wind speed alone is not an indicative which cross section is more critical. 
It can be decided after a complete stress analysis considering the probability of occurrence of the wind speeds and the number of oscillation cycles in the design life of the structure which is not in the scope of this study. 

Buffeting analysis makes use of the static wind coefficients, aerodynamic derivatives and aerodynamics admittance functions to compute aerodynamic forces on the section. 
Aerodynamic derivatives are associated to the motion-induced forces. 
The shape of the section different from a flat plate means that we are moving apart from the flat plate assumptions here  and the sears admittance can not be used. 
Therefore aerodynamic admittance functions have been computed for all the section geometries.
The results of the buffeting analysis are presented in Figure~\ref{Figure:FlutterVIVBuffeting}. 
The buffeting forces tend to reduce for a shallower and wider section. 
The largest buffeting forces are experienced by narrow and deep sections. 
This is due to the fact that the sections will have low stiffness in that direction. 
Th largest rotation is seen for a shallow and narrow section due to its low torsional stiffness. 
The buffeting forces are used for the stress checks and govern the plate thicknesses. 
However, the plate thicknesses have not been considered as variable in this study. 
The stresses have been checked for buffeting forces along with the dead and superimposed dead. 
The stresses in the deck and cables have been checked against the allowable values. 
Again the deeper and narrower cross section is more critical for the overall stresses.

\subsection{Aerodynamic optimization}
\label{Section:AerodynamicOptimization}
The optimization process minimizes an objective while satisfying certain constraints. 
Here the mass of the deck is required to be minimized. 
For this purpose, the objective function is set as minimize the total mass of the deck for the whole span length $L$. 

The constraints are summarized in Table~\ref{Table:Constraints}. 
The maximum vertical deck displacement $h_{max}$ under buffeting wind actions is checked which must be less than the maximum limit $P_{B,lim}$. 
On the same lines, the maximum allowed lateral displacements $p_{max}$ and rotations $\alpha_{max}$ of the deck have been defined in the constraints. 
These limits are checked at each node along the span. 
In addition to that, stress limit in the deck $\sigma^{deck}_{B,lim}$ is considered. 
Main cable tensile stress limit $\sigma^{cable}_{B,lim}$ is also taken into account. 
Flutter limit is checked against the maximum allowed design wind speed at the site with some safety factor $U_{F,lim}$. 
The VIV amplitudes of displacements are required to be examined. 
The peak resonant vertical displacements in the deck are compared with the limiting values $h_{V,lim}$. 
A total of 927 constraints have been checked for each sample. 

The cost function is defined here as 
\begin{equation}
	\label{Equation:CostFunction}
	\textrm{cost}=\frac{m(d_x,b_x)}{m_{avg}},
\end{equation}
where $m$ is the section mass per unit length and $m_{avg}$ is the average mass of the sections per unit length in the sample set. 
The denominator $m_{avg}$ used here is just to normalize the cost function. 
It is to be noted that the section is same throughout the length of the bridge, therefore $L$ is not included in the equation. 
Eq.~\eqref{Equation:CostFunction} shows that the wight of the section 
is to be minimized in the optimization algorithm provided the constraints in Table~\ref{Table:Constraints} are satisfied.

\begin{table*}[h]
	\caption{Constraints considered in the aerodynamic optimization process (cf.~Eq.~\eqref{Equation:GeneralConstraints}).}
	\label{Table:Constraints}
	\centering
	\begin{tabularx}{\textwidth}{C{0.3}C{0.3}C{0.5}C{0.8}L{1.7}C{0.5}L{2.0}L{0.3}}
		\hline
		$i$ & Case 	& $P$ 					& $P_{max}, P_{min}$		& Limiting Criteria & Locs.	& Description & $s$\\
		\hline
		\multicolumn{6}{l}{Buffeting}\\
		1 & I 	& $h_{max}$ 			& $h_{B,lim}$ 				& $L_{span}/250=2400$ mm	& 93 	& no. of nodes & 93\\
		2 & I 	& $p_{max}$ 			& $p_{B,lim}$ 				& $L_{span}/650=925$ mm		& 93 	& no. of nodes & 93\\
		3 & I 	& $\alpha_{max}$		& $\alpha_{B,lim}$ 			& $1^\circ$ 				& 93 	& no. of nodes & 93\\
		4 & I 	& $\sigma^{deck}_{max}$ & $\sigma^{deck}_{B,lim}$	& $250$ MPa 				& 90 	& no. of beam elements & 90\\
		5 & I 	& $\sigma^{cable}_{max}$ & $\sigma^{cable}_{B,lim}$	& $800$ MPa 				& 184 	& no. of cable elements & 184\\
		\hline
		\multicolumn{6}{l}{Vortex-induced vibrations}\\
		6 & I 	& $h_{res}$ 			& $h_{V,lim}$ 				& $0.04/f_h$				& 93 	& vertical modes $\times$ nodes & 372\\
		\hline
		\multicolumn{6}{l}{Flutter}\\
		7 & II 	& $U_{cr}$ 				& $U_{F,lim}$ 				& $1.20\times U_m=75$ m/s	& 1 	& SDOF and 2DOF flutter & 2\\
		\hline
	\end{tabularx}
\end{table*}

\begin{figure}[!htbp]
	\centering
	\includegraphics[scale=0.9,trim=0mm 6mm 0mm 10mm,clip]{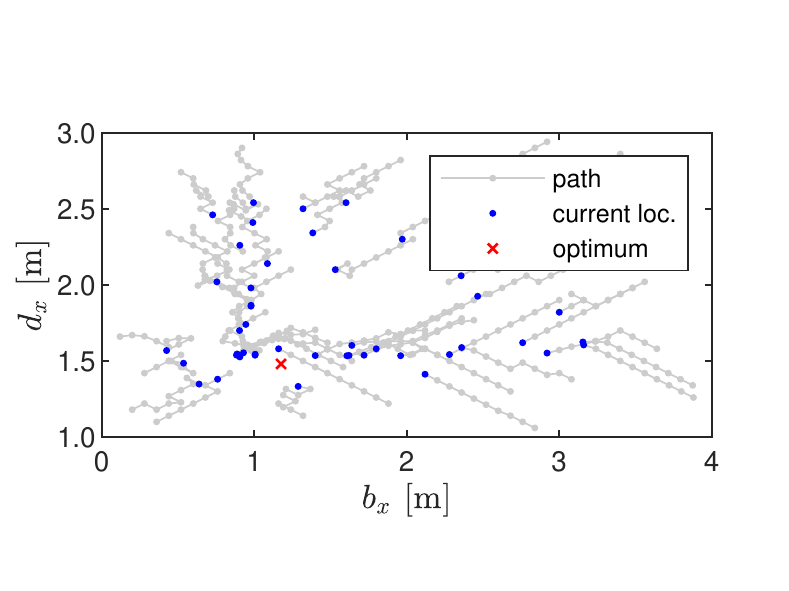}\\
	\includegraphics[scale=0.9,trim=0mm -1mm 0mm 2mm,clip]{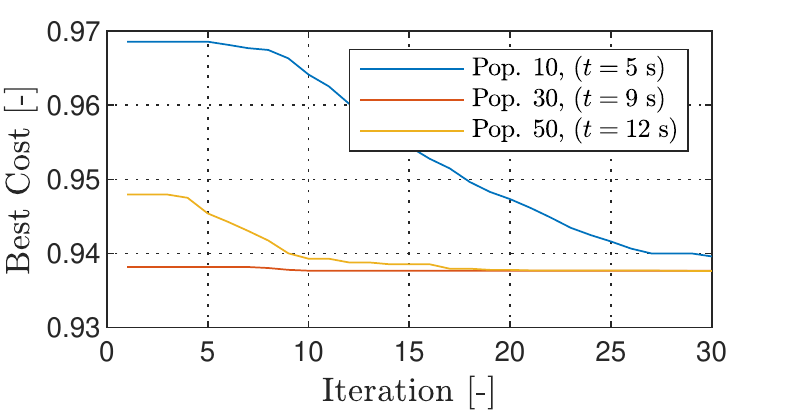}\\
	\caption{Optimization using PSO algorithm: (left) instantaneous view of the particle population at iteration number 10, (right) sensitivity of the cost function  based on population size.}
	\label{Figure:Optimization}
\end{figure}

The particle swarm optimization algorithm presented in Section~\ref{Section:OptimizationStrategy} has been utilized here. 
The optimization scheme is implemented in the MATLAB environment. 
The optimization parameters and required input is set in the script. 
The algorithm generates samples in the first iteration randomly for the section geometry and computes the target quantities of interest. 
For this purpose, response surfaces developed in Section~\ref{Section:Response surface generation} have been used. 
The response surface takes the sampled values in the form of $d_x$ and $b_x$ and produces output quantities from the aeroelastic analyses. 
The predicted quantities for each sampled section geometry are then checked against the constrains if satisfied or not. 
The samples which are not satisfied are discarded and the remaining samples are considered to calculate the cost. 
The process is repeated for the each iteration. 
Since a response surface has been utilized, the computation for each iteration takes less than a second for the whole population. 
Figure~\ref{Figure:Optimization} shows an instantaneous plot of the optimization process. 
The particles randomly generated initially are moving towards a global minima indicated as `optimum' and the traced paths are shown. 
The movement in the traced path is also not straight as there is randomness in the location  of the particle themselves to explore the localized regions. 
The number of iterations were chosen as 30 beyond which there was not a significant improvement. 
The optimization process was also repeated 6 times each by generating a new initial sample population to make sure that the algorithm is not trapped in a local minima. 

The figure also shows the progress of the algorithm to compute the cost of the `global best'. A sensitivity study has been conducted by selecting different population size. 
For this purpose, a population size of 10, 20 and 50 was considered. 
The plot indicates that all cases approached towards the global minima; however, the population size 10 was unable to capture the lowest cost and may require much more iterations. 
This is due to the fact that the algorithm is trapped in a local minima. 
Here randomness introduced may help to come out of this local minima which may need a few more iterations but it is not guaranteed. 
The time required between different population sizes is not that large indicating a larger population set can be considered. 
The population size of 30 and 50 converged much earlier and surprisingly for case 30 performed even better than case 50. 
This is because the initial generation of the population is independent and random. 
If the generated population ends up very close to the global minima, this will be the advantage right from the beginning. 
Nevertheless, this sensitivity study indicates that the method is stable in finding the global minima and a reasonable population size is able to provide the solution in few iterations. 

It may require to do a sensitivity study to find a reasonable population size. 
A smaller population may be insufficient to cover a wider region accurately while being unable to explore the localized regions. 
This will make make the algorithm to be trapped in a local mimima. 
A very large population size on the other hand will make the optimization problem inefficient. 
However, it may have better chances of finding global minima.

\begin{figure}[!htbp]
	\centering
	\includegraphics[scale=0.9,trim=4mm 0mm 8mm 0mm,clip]{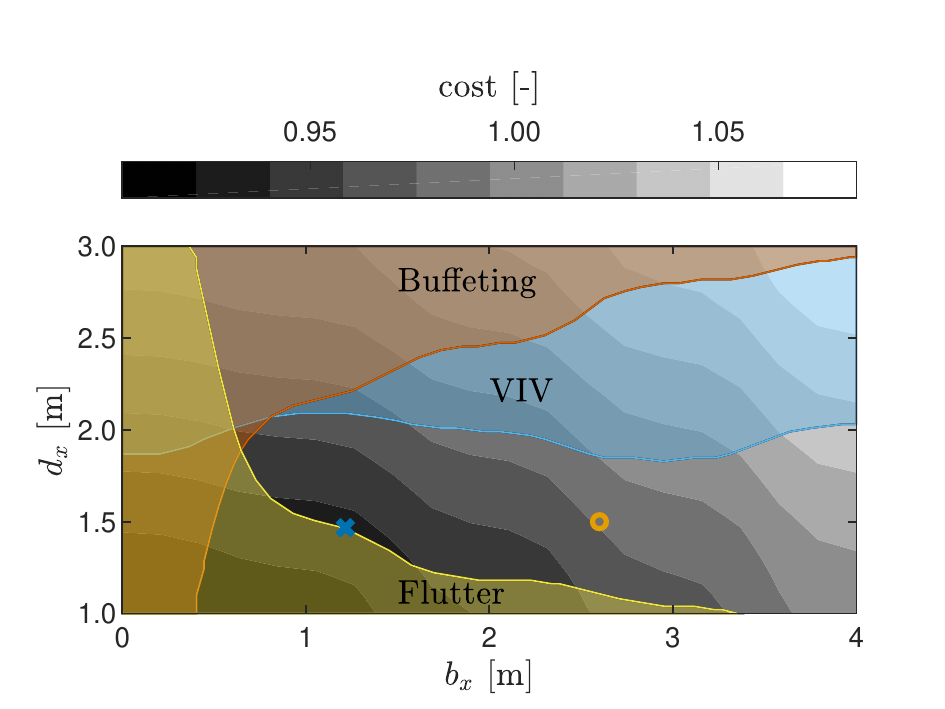}
	\caption{Cost function of the optimization process and the constraints. Red cross indicates the optimized solution from PSO. (circle) Reference section: $b_x=2.600$, $d_x=1.500$, cost$=0.987$ (cross) Optimum solution: $b_x^*=1.176$, $d_x^*=1.480$, cost$=0.938$}
	\label{Figure:OptimizationSurface}
\end{figure}

Figure~\ref{Figure:OptimizationSurface} depicts the overall optimization results in terms of cost and the limits set for different aeroelastic phenomena. 
It is clear from the figure that the deeper sections perform poorly in case of VIV. 
The deeper sections have large lift forces due to vortex shedding and have low Strouhal number. 
Here both the classical flutter and torsional flutter responses have been checked for the section geometries.
Classical flutter is critical for narrow and flat sections whereas rectangular deep sections are prone to torsional flutter. 
Buffeting response has been checked for displacements and stresses. 
Buffeting is seen to be critical for narrow and deeper sections. 

\begin{figure}[!htbp]
	\centering
	\includegraphics[scale=0.9,trim=10mm 40mm 7mm 25mm,clip]{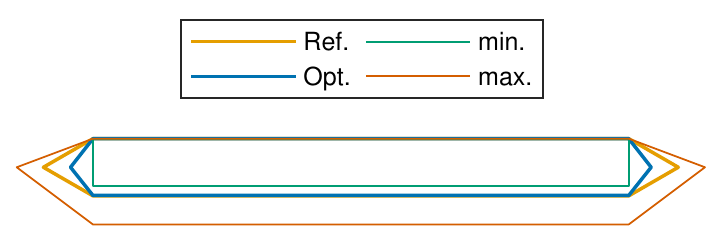}
	\caption{Cross section geometry comparison with the optimized shape. (Optimum solution: $b_x=1.176$, $d_x=1.480$)}
	\label{Figure:GeometryLimits}
\end{figure}

Figure~\ref{Figure:GeometryLimits} shows the comparison of the optimized shape with the reference shape. 
Also the maximum and minimum limits considered in the analyses have been shown. 
It is observed that the the optimized shape is not significantly different form the original reference section. 
Only a smaller fairings can be sufficient in this case. 
This also reinforces the argument that the original section was designed on the basis of several Wind tunnel rests abut this would be made economical by performing optimizations study as presented in this paper. 
The weight of the reference section is 11.68~t/m and the optimized section has a reduction in weight of 4.71\%. 
It may seem not a very significant improvement. 
However, the aim was not to to improve the existing cross section rather the idea here is to show that the framework can be applied to such scenarios to optimize the cross section at the design stage.

It is important to note that the structural parameters such as thickness of the deck plates and cable sizes have not been considered as a variable in the optimization problem. 
If that is the case then the optimization algorithm must include these variables and the response surface needs to be updated. 
This will make use of the existing aerodynamic properties already determined from the CFD simulations. 
The aeroelastic analyses have to be performed considering these changes. 
It may require more samples to build a response surface of reasonable quality. 
But it would be beneficial at the optimization stage. 
Since the optimization loop is outside the aeroelastic analysis and response surface, the optimization process which requires a large population set and several iterations would become significantly fast. 

With a population size of 50 and 30 iterations a total of 1500 analyses are required whereas for a population size 30 reduces it to 900. 
This is quite a significant number but just 25 number of samples have been used to build the response surface. 
In contrast to gradient based optimization algorithms where a global minima is not guaranteed the number of required analyses are large but the optimization time is reduced to a few seconds. 

The optimized geometry is also very much dependent on the constraints. 
The constraints can change and a different geometry is achieved. 
This can be seen in case of probabilistic constraints where for each constraint is defined with a certain probability. 
However, in this study it is not considered. 


\section{Conclusion}
\label{Section:Conclusion}
This paper presents the framework to perform aerodynamic shape optimization of a bridge deck by utilizing CFD simulations and response surface strategies. 
Here the focus is on the influences section geometry variation on the aeroelastic response and how it can be utilized to minimize the section weight considering the relevant limit states. 

The framework is structured in two stages. 
The first stage is the development of a response surface. 
The cross section of the bridge is parametrized using width and depth variables. 
The variation of these geometry parameters allows a wide range of cross section geometries. 
The samples are generated for the section geometries for which the subsequent analyses have to be performed. 
The structural properties are determined from a fine element model of the bridge. 
The CFD simulations on sampled geometries in the design domain are carried out to obtain the aerodynamic parameters. 
The governing aeroelastic analyses are performed. 
These include flutter limits computed from eigenvalue analysis, VIV response based on Eurocode and buffeting response following linear unsteady theory considering static wind coefficients, aerodynamic derivatives and aerodynamic admittance functions. 
The geometry parameters and the output response quantities are utilized to develop response surfaces using moving least squares (MLS). 
MLS performs better than the polynomial regressions as it considers the localized region near the point computed. 
These response surfaces provide the output quantities from the aeroelastic analyses considered in the scheme. 

In the next stage, optimization was performed by non-gradient based algorithms.
Here, particle swarm optimization (PSO) algorithm has been adopted considering geometric constraints and limit states. 
The main objective is to minimize the weight of the deck section by satisfying the essential criteria for all the relevant aeroelastic phenomena. 
For this purpose, the response surfaces developed in the first stage have been used. 
The optimization algorithm requires a large number of samples which makes the process very time consuming. 
Here the response surface approach makes this process significantly efficient and it can be completed in a few seconds. 
The response surface required just 25 sample geometries whereas the optimization process required up to 1500 samples. 
Moreover there are several steps involved in the process after which a section geometry is judged. 
Running an optimization scheme without using response surfaces would make it not only inefficient but it will not be robust. 
As any flaw in the sequence will break the algorithm and the whole cycle will need to be started from the beginning. 
This is especially important considering the chain of steps taken for the one complete analysis. 
Especially when the analysis is to be repeated for a sensitivity due to parameters of optimization algorithm. 

The application of the developed scheme has been presented for a suspension bridge. 
Lilleb\ae lt suspension bridge has been considered as a reference structure. 
Although the optimized cross section geometry is not significantly different from the original section geometry of the bridge, the effectiveness of the approach quite evident. 
The optimized section is 4.71\% lighter than the reference section which is till an improvement. 

The sensitivities due to different input parameters are also discussed. 
It has been found that the considered aeroelstic phenomena limits cover different area of the design space. 
Generally, deeper sections are prone to VIV response. 
Buffeting is governing criterion for narrow and deep sections. 
Whereas flutter is critical for shallow and narrow sections. 

The paper provides a comprehensive framework to include several aeroelastic phenomena for aerodynamic shape optimization which has the potential to be used in achieving a high performance aerodynamic shape for the long-span bridges. 
The framework has been found computationally very efficient in terms of solving such a large scale problem. 
The structural optimization along with aerodynamic optimization can be investigated in future the works. 
The optimization can also include fatigue analysis for buffeting and VIV response.




\section*{Acknowledgements}
This research is supported by the German Research Foundation (DFG) [Project No. 329120866], which is gratefully acknowledged by the authors.

\bibliographystyle{myunsrtnat}
\setlength{\bibsep}{1.2pt}
\small{\bibliography{MyBib}}

\end{document}